\documentclass[sigconf]{acmart}
\AtBeginDocument{%
  }

\setcopyright{acmlicensed}
\copyrightyear{2026}
\acmYear{2026}
\setcopyright{cc}
\setcctype{by-nc-nd}
\acmConference[CHI '26]{Proceedings of the 2026 CHI Conference on Human Factors in Computing Systems}{April 13--17, 2026}{Barcelona, Spain}
\acmBooktitle{Proceedings of the 2026 CHI Conference on Human Factors in Computing Systems (CHI '26), April 13--17, 2026, Barcelona, Spain}
\acmPrice{}
\acmDOI{10.1145/3772318.3790866}
\acmISBN{979-8-4007-2278-3/2026/04}

\usepackage{array}
\usepackage{subcaption}
\usepackage{xcolor}
\usepackage{enumitem}
\usepackage{tabularx}
\usepackage{float}

\usepackage{listings}

\lstdefinestyle{promptstyle}{
    basicstyle=\ttfamily\footnotesize, 
    breaklines=true,                   
    breakatwhitespace=true,            
    frame=single,                      
    rulecolor=\color{black!30},        
    backgroundcolor=\color{gray!5},    
    showstringspaces=false,            
    captionpos=b,                      
    columns=fullflexible,              
    keepspaces=true                    
}

\newcommand{\system}{InterFlow}

\newcommand{\h}[1]{{#1}}

\begin{document}

\title{\system{}: Designing Unobtrusive AI to Empower Interviewers in Semi-Structured Interviews}

\author{Yi Wen}
\affiliation{%
  \institution{Texas A\&M University}
  \city{College Station}
  \country{United States}
}
\email{cyberwenyi2357@tmau.edu}

\author{Yu Zhang}
\affiliation{%
  \institution{City University of Hong Kong}
  \city{Hong Kong}
  \country{Hong Kong}
}
\email{yui.zhang@cityu.edu.hk}

\author{Sriram Suresh}
\affiliation{%
  \institution{Texas A\&M University}
  \city{College Station}
  \country{United States}
}
\email{srisuresh@tamu.edu}

\author{Zhicong Lu}
\affiliation{%
  \institution{George Mason University}
  \city{Fairfax}
  \country{United States}
}
\email{zlu6@gmu.edu}

\author{Can Liu}
\affiliation{%
  \institution{City University of Hong Kong}
  \city{Hong Kong}
  \country{Hong Kong}
}
\email{canliu@cityu.edu.hk}

\author{Meng Xia}
\authornote{Corresponding author}
\affiliation{%
  \institution{Texas A\&M University}
  \city{College Station}
  \country{United States}
}
\email{mengxia@tamu.edu}

\renewcommand{\shortauthors}{Wen et al.}

\begin{abstract}

Semi-structured interviews are a common method in qualitative research. However, conducting high-quality interviews is cognitively demanding and requires strong interviewing skills. To lower this bar, we propose InterFlow, an AI-powered visual scaffold that helps interviewers manage the interview flow and facilitates real-time data sensemaking. The system dynamically adapts the interview script to the ongoing conversation and provides a visual timer to track interview progress and conversational balance. It further supports information capture with three levels of automation: manual entry, AI-assisted summary with user-specified focus, and a co-interview agent that proactively surfaces potential follow-up points. A within-subject user study (N = 12) indicates that InterFlow reduces interviewers’ cognitive load and facilitates the interview process. Based on the user study findings, we provide design implications for unobtrusive and agency-preserving AI assistance under time-sensitive and cognitively-demanding situations.

\end{abstract}



\begin{teaserfigure}
    \centering
\includegraphics[width=0.85\linewidth]{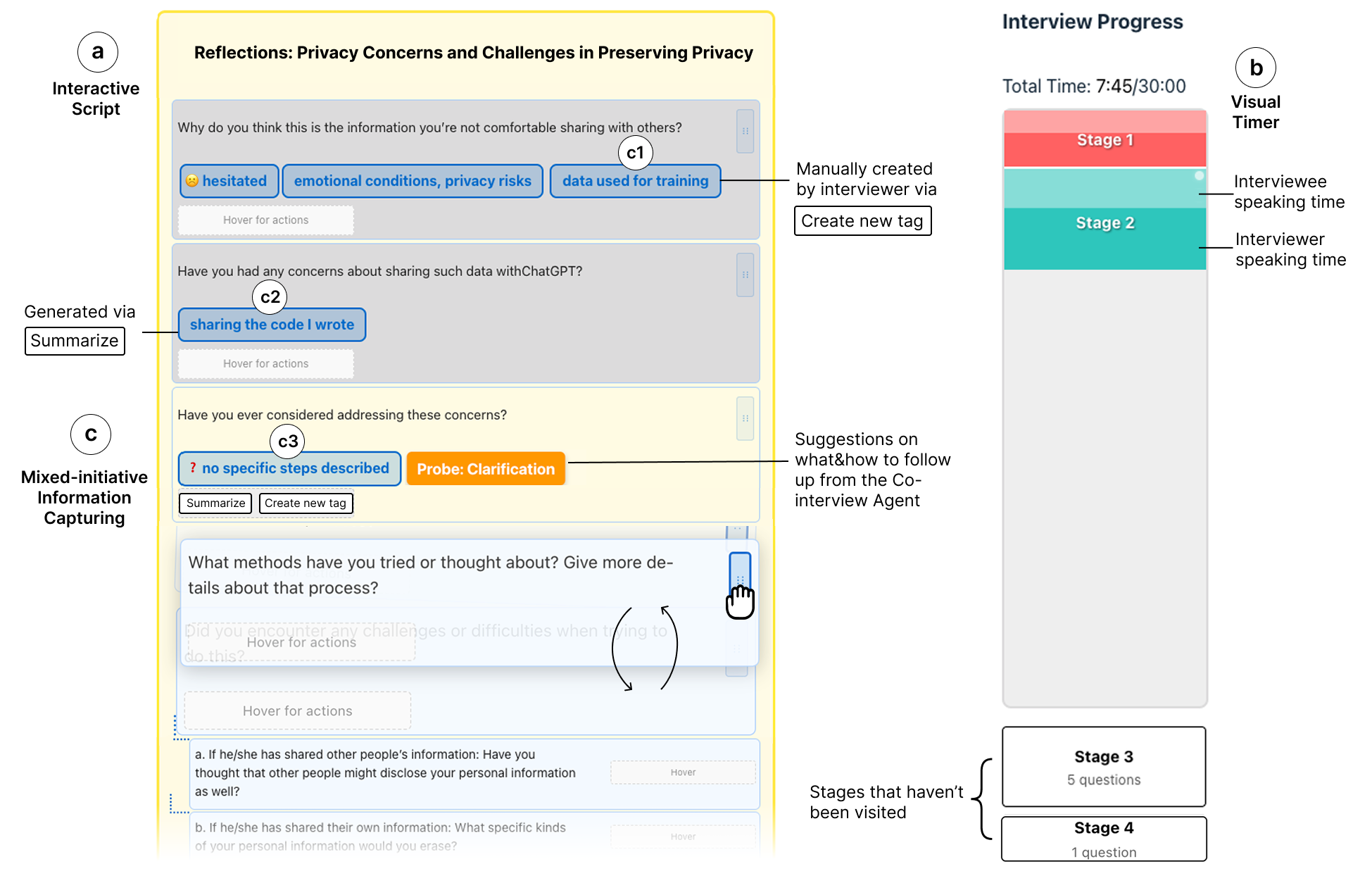}
    \caption{Main user interface of \system{}, an AI-powered visual scaffold for conducting semi-structured interview. It consists of (a) an interactive script that dynamically visualizes the structure and status of the interview (the ongoing question in yellow, visited questions in gray, and unvisited ones in blue), and allows interviewers to adjust the question sequence via drag-and-drop; (b) a visual timer that provides an overview of the interview progress and conversational balance; and (c) a mixed-initiative information capture scheme with three levels of automation: (c1) manual entry (click on "Create new tag"), (c2) instant summary with user-specified focus (click on "Summarize"), and (c3) a Co-Interview agent that proactively suggests what to follow up on and how to do so. }
    \Description{Screenshot of InterFlow, an AI-assisted semi-structured interview interface. On the left is an interactive interview script with questions color-coded by status (current, visited, and unvisited) and draggable to change order. On the right are a visual timer showing interview progress and speaking balance, and a panel where interviewers can either manually add tags, request focused summaries, or view AI-suggested follow-up questions.}
    \label{fig:teaserFigure}
\end{teaserfigure}

\maketitle

\section{Introduction}

Semi-structured interviews are a common method in qualitative research, where the interviewer follows a predefined set of questions, while also asking follow-up questions to seek in-depth responses~\cite{adams2015conducting}. It is used for exploring complex social phenomena or obtaining participants' subjective experiences and attitudes, widely adopted in multiple disciplines such as Anthropology~\cite{bernard2012research}, Sociology~\cite{brinkmann2014unstructured}, and Human Computer Interaction (HCI)~\cite{blandford2016qualitative}. A complete interview process generally involves preparing for, setting up, conducting, and analyzing the interviews; while the quality of the interview itself is especially important, as it directly impacts the effectiveness of later analysis, verification, and reporting~\cite{adams2015conducting}. 

Obtaining rich, goal-relevant data in a semi-structured interview is more challenging than it appears, as the task places a high cognitive load on interviewers, who need to simultaneously listen actively, formulate pertinent follow-up questions, and sustain a natural conversational flow \cite{doingInterviews,fontana2000interview,kvale2009interviews}. Mastering the craft requires personal judgment and practical skills that largely come from experience \cite{gesch2015reflecting}. Despite the existence of numerous handbooks that provide guidance to interviews \cite{Boekhoven2024ListeningTP,hermanowicz2002great,fontana2000interview}, many interviewers, especially non-experts who have not received formal training and are less experienced~\cite{gesch2015reflecting}, find it challenging to translate these theoretical guidelines into effective practice during actual interviews~\cite{challengesWhenConductingInterview, workingThroughChallenges, pope1974experienced,joysandchallenges}. Specifically, non-expert interviewers have difficulties in maintaining a logical progression of questions, being flexible with the interview script,  appropriately pacing the conversation, probing effectively, and monitoring data quality \cite{gesch2015reflecting,roulston2003learning,dejonckheere2019semistructured}. 

Recent works in Natural Language Processing (NLP) \cite{meng2023followupqg,wong2025ai} and HCI ~\cite{hu2024designing,xiao2020if} have explored ways to automate the qualitative data collection process by incorporating humans' high-order thinking skills and knowledge into Language Models, enabling them to generate contextually appropriate questions in response to interviewees’ answers. However, empirical comparisons showed that while AI (here referring to LLM-powered conversational agents) is powerful in understanding natural language conversations and generating appropriate questions, it still cannot outperform human interviewers~\cite{hu2024designing}, largely due to a lack of grounding~\cite{clark1991grounding}. LLMs exhibit limited use of acknowledgements or affirmations to foster comfort and trust, struggle to recognize when questions have been adequately answered, and lack the ability to strategically steer interviews through multi-turn planning. ~\cite{lu2024newsinterviewdatasetplaygroundevaluate}. A recent study~\cite{liu2025envisioning} also showed that interviewers expect AI to serve as a supportive tool working alongside them, but not replacing them or taking the lead.

In this sense, we aim to design a tool that brings together the complementary strengths of both human and AI, bridging theoretical guidelines and interviewers' real-world practices. However, integrating AI assistance in real time presents design challenges, as it may distract interviewers from the conversation with interviewees, and interviewers often have limited cognitive bandwidth and time to reflect on AI-generated suggestions, which can lead to inappropriate reliance on the system \cite{cao2023time}. To better understand the challenges faced by non-expert interviewers and to inform a design that is both feasible for real-time use and compatible with interviewers’ existing workflows, we conducted an observational study with six non-expert interviewers. We found that they struggle to capture information without disrupting conversational flow, often wishing for a partner to assist with note-taking and supplemental questions. They also face difficulties navigating scripts and balancing the depth and breadth of the interview. From these formative insights, we derived two design goals: (1) helping interviewers manage interview flow by supporting script navigation and situational awareness, and (2) facilitating real-time data sensemaking by reducing the effort required for note-taking and surfacing noteworthy follow-up points.

We propose \system{}, an intelligent visual scaffold that enables the static interview script to evolve with the ongoing conversation. To help interviewers manage the interview flow, \system{} takes the pure-text script and transforms it into interactive UI components that reflect the structure of the interview, the ongoing question, and the time spent on each stage, along with the interviewer's speaking ratio. To facilitate real-time data sensemaking, it supports mixed-initiative information capture that allows users to take notes manually, let AI summarize the recent conversation, and receive follow-up suggestions from a proactive “co-interviewer” agent.

\h{We evaluated the system with a within-subject study (N=12), comparing \system{} with a baseline system which consists of a text editor and an AI assistant powered by OpenAI's speech AI model \footnote{\url{https://platform.openai.com/docs/guides/realtime}}. Results showed that \system{} reduces the cognitive load of doing interviews, helping interviewers maintain an overall picture of the interview while keeping a record of the details they considered important. 
The system’s multi-level, process-centered AI suggestions impose minimal cognitive overhead and fit naturally into interviewers’ thinking flow. However, the perceived utility of the system's suggestions varies across users and the actionability of these suggestions — namely, whether interviewers are able to turn them into timely follow-up questions — can be constrained by the rapidly evolving conversational dynamics of real-world interviews.}
Based on the findings, we further generalize the design implications into real-time AI assistance in similar settings where users are primarily engaged in a attention-intensive task and decisions must be made under time pressure.

Overall, our work offers the following contributions:
\begin{itemize}[leftmargin=*, itemsep=5pt, topsep=5pt]
\item{An observational study that reveals the non-expert interviewers' real-world practices and challenges of conducting semi-structured interviews}
\item{The design and implementation of \system{}, an intelligent visual scaffold that provides unobtrusive assistance for conducting semi-structured interviews}
\h{\item{A user evaluation demonstrating the usability, usefulness and drawbacks of \system{}, and implications for designing unobtrusive AI assistance that can fit into user's thinking flow under time-sensitive and cognitively-demanding situations}}
\end{itemize}

\section{Related Work}

\subsection{AI Support in Qualitative Research}
In recent years, there has been a growing interest in exploring how AI can support qualitative research \cite{jiang2021supporting,gao2024collabcoder,sinha2024role,schroeder2025large,gao2023coaicoderexamining}. Schroeder et al.~\cite{schroeder2025large} summarized the broad range of uses of large language models (LLMs) in data collection, data analysis, ideation, writing, and finding related work. Most existing research has concentrated on the task of qualitative analysis, with an emphasis on how to preserve researcher agency while still being supportive, positioning AI as an assistive role to foster practices of responsible science ~\cite{birhane2023science}.

By contrast, prior work on the semi-structured interview process itself has largely focused on automating data collection, for instance, through conversational agents ~\cite{xiao2020if,su2019follow,hu2024designing} or chatbots~\cite{xiao2020tell,tallyn2018ethnobot}. While these approaches demonstrate the potential of AI in information elicitation, they often target less flexible contexts such as job interviews or ethnographic data gathering, and their performance in more open-ended interviews has been limited ~\cite{lu2024newsinterviewdatasetplaygroundevaluate}. Most recently, Liu et al. \cite{liu2025envisioning} explored how researchers envision AI assistance in semi-structured interviews, surfacing different needs of interviewers of various expertise levels, and design challenges related to time constraints and the triangular dynamic among the interviewer, interviewee, and AI-driven assistant.
\h{Complementary to this, Li et al. \cite{li2025insightbridge} designed a medium that enhances empathy and shared meaning-making between the interviewer and interviewee, addressing the challenges rooted in interpersonal communication. From this work, we see the feasibility of providing real-time support with interview. However, our work concentrates on the procedural and cognitive challenges interviewers face, such as managing the interview flow and identifying follow-up opportunities.}

On top of previous work, we aim to help non-expert interviewers, who constitute a substantial portion of the broader researcher community, adhere to theoretical guidelines that can potentially lead to high-quality interviews. To our best knowledge, no one has built tools for addressing this problem before.

\subsection{Designing Unobtrusive AI for Attention-Intensive Tasks}

Previous work has explored how AI systems can be introduced into contexts where users must sustain high levels of attention in a main activity, such as clinical decision making \cite{unremarkable,lorenzini2023artificial}, remote meetings\cite{areWeOnTrack,meetmap,meetingCoach,chen2023meetscript,chandrasegaran2019talktraces,son2023okay,coco, zhang2025speechcap}, and classroom settings \cite{tang2024vizgroup,glancee,sato2023groupnamics}.
Across these domains, AI assistance has been shown to enhance performance by providing additional perspectives \cite{unremarkable}, surfacing overlooked issues\cite{areWeOnTrack}, and supporting sensemaking and timely reflection \cite{tang2024vizgroup}.

However, as prior studies suggest, AI support under time pressure carries risks of increasing users' cognitive load ~\cite{cao2023time} and inappropriate reliance~\cite{swaroop2024accuracy} on AI, making it essential to carefully design such systems to avoid distracting users from their primary tasks and to mitigate under- or over-reliance on AI. Effective strategies for avoiding distractions include embedding AI outputs into existing workflows rather than requiring explicit interaction \cite{unremarkable}, keeping information presentation lightweight and ambient (e.g., sidebars, icons, or subtle highlights) to support quick glances rather than continuous attention \cite{glancee,tang2024vizgroup,chandrasegaran2019talktraces,areWeOnTrack}, and timing interventions so they appear only during natural pauses or at moments of genuine need \cite{areWeOnTrack,unremarkable}. Some prior work enables users to configure system behaviors in advance to provide more personalized assistance to different users\cite{tang2024vizgroup,areWeOnTrack}. In explainable AI field, methods like cognitive forcing ~\cite{buccinca2021trust} and presenting human correctness likelihood ~\cite{ma2023should} have been proposed to promote appropriate trust in AI-assisted decision making.

However, the work and methods introduced above largely focus on structured tasks with clear rules and routines, whereas semi-structured interviews are open-ended and exploratory in nature, making existing strategies difficult to directly apply in the context of semi-structured interviews.

\section{Formative Study}
\subsection{Background and Purpose}
Foundational works in qualitative methodology~\cite{doingInterviews} define high-quality interviews by the richness and relevance of responses and the degree of clarification achieved. To support such quality, domain experts recommend practices such as building rapport, employing active listening, probing for deeper meaning, maintaining neutrality, and documenting non-verbal cues and follow-up questions~\cite{hermanowicz2002great}. However, recent findings suggest that interviewers are more receptive to technology when it assists with cognitive rather than emotional tasks, as technological involvement in sensitive matters may harm interviewees’ well-being~\cite{liu2025envisioning}. Motivated by this, we focus on the fundamental cognitive challenge of managing the interview flow~\cite{gesch2015reflecting}.

Our attention is on non-expert interviewers, who represent a large portion of practitioners and often struggle with conducting semi-structured interviews effectively. Prior studies highlight their mistakes and difficulties~\cite{gesch2015reflecting,pope1974experienced}, yet little is known about how they operate in real-world settings, the specific reasons behind their challenges, the tools they rely on, and how AI might fit into their workflow. Moreover, designing tools for real-time use in semi-structured interviews demands special care due to the high cognitive load such interviews impose~\cite{liu2025envisioning}. To address these gaps, we conducted a formative study with non-expert interviewers.

\subsection{Participants and Procedure}
\h{We conducted an observational study followed by an interview with six participants who have basic knowledge and experience with qualitative interviews but do not consider themselves as experts. We recruited them through known contacts and snowball sampling. Similar to prior work on supporting novices in writing revision \cite{friction}, we asked them to report their past experience with semi-structured interviews and self-rate their expertise level from 1 (beginner interviewer) to 7 (expert interviewer). Since it is difficult for participants to recall the exact number of interviews they have conducted, we asked them to select from a set of predefined ranges $\,(0,5],\ (5,10],\ (10,15],\ (15,20],\ (20,\infty)$. The average expertise level of the participants is 3.33 (SD=1.21). 1/6 participant had 0–5 interviews in the past, 2/6 had 5–10, 2/6 had 10–15, and 1/6 had 15–20. The detailed demographic information of the participants is included in \autoref{tab:observational-demographics} in Appendix.}

In the observational study part, the first author sat in the interview session of the participant who was conducting a semi-structured interview for their own research project. The first author took notes on the key phenomena observed during the interview. After the session ended, the first author asked the participants about the reasons behind their behaviors and the shortcomings of the current tools they use.
\h{\subsection{Analysis}
For ethical reasons, we did not record the participants’ original interview sessions, which involved their own study participants. We only recorded the audio of the follow-up interviews between the first author and the participants. We conducted a thematic analysis \cite{terry2017thematic} of the observational notes and interview transcripts. Following an inductive coding \cite{Chandra2019} approach, the first author conducted line-by-line coding to identify the common practices, mistakes and challenges of conducting interviews. Codes were iteratively grouped into higher-level categories, and the research team met regularly to discuss interpretations and refine emerging themes. The findings of our formative study are presented below.
}

\subsection{Findings}

\subsubsection{\textbf{[F1]} Difficulties with navigating the interview script}
It is common for the interview not to proceed as planned, so the interviewer needs to adjust the script on the spot, such as changing the order of questions or skipping a question~\cite{adams2015conducting,kvale2009interviews}.
However, we observed that the participants seldom do so. When asked about the reason for that, P6 explained ``A lot of times I tried to take the conversation on a natural route, so if the next question that needs to be asked is not at all related to what the person just talked about, I'll skip over it. However, I may end up forgetting to ask that question because of the adjustment.''  The opinion of unwillingness to take the risks of adjusting the script is echoed by P4 ``I feel very stressed already, I'm afraid I don't have extra cognitive bandwidth for remembering the changes I made.'' Beyond adjustments, interviewers also expressed difficulties in locating their current position within the script and keeping track of which questions had already been covered or remained, which further contributed to their reluctance to deviate from the script. ``It sounds trivial, but it can be hard for a green hand like me,'' said P2. 

\subsubsection{\textbf{[F2]} Interrupted conversation due to manual note-taking} 
Conducting interviews poses a high demand on working memory, as new information keeps emerging during the conversation. Based on the observation, most participants (5 out of 6) take notes during the interviews. They typically write down respondents' answers as well as their emerging thoughts or insights to avoid repetitive questions, assist their understanding and interpretation of interviewees' responses, help decide if a follow-up question is needed, and deepen impressions for more efficient analysis later. However, we observed that manual note-taking is time-consuming, and sometimes the interviewers have to pause the conversation to take notes.  As P5 explained, ``I know it's not good, but I just can't help doing so since I feel nervous about forgetting a follow-up point that needs to be put forward later if I don't write it down.'' 

\subsubsection{\textbf{[F3]} Lack of awareness of time allocation} 
Although the participants have the guidelines (e.g., do not steer the conversation but listen to the interviewee, balance topic coverage and conversational depth) in their mind, sometimes they still unintentionally talk too much,  spend too much time on one topic, or the interview exceeds the planned time.
Without external tools that record and report their time usage, it's challenging for them to manage their time allocation effectively. 
Participants expressed their needs about the reminder: ``Just having a subtle prompt about how much time has elapsed, how much time is left, how much time you've been spending on this certain theme, '' said P4. 

\subsubsection{\textbf{[F4]} Collaborative interviewing: two interviewers for workload sharing } 
Among the six interview sessions we observed, one session had two interviewers collaboratively working on the interview. One is the main interviewer who asks most of the questions; the other one just listens to the conversation and takes notes. After the main interviewer finishes asking, the other interviewer will ask a few follow-up questions to delve deeper into certain areas. When being asked about this ``co-interview'' setting, all participants feel that it would make them more relaxed. ``There are many things that I need to take care of during interviews. Having someone to share the tasks with me is desirable and usually leads to better interview quality,'' noted P2. ``However, it's hard to find someone who also understands the project well, and time coordination is another difficulty.''

\begin{figure*}[!ht]
    \centering
    \includegraphics[width=\linewidth]{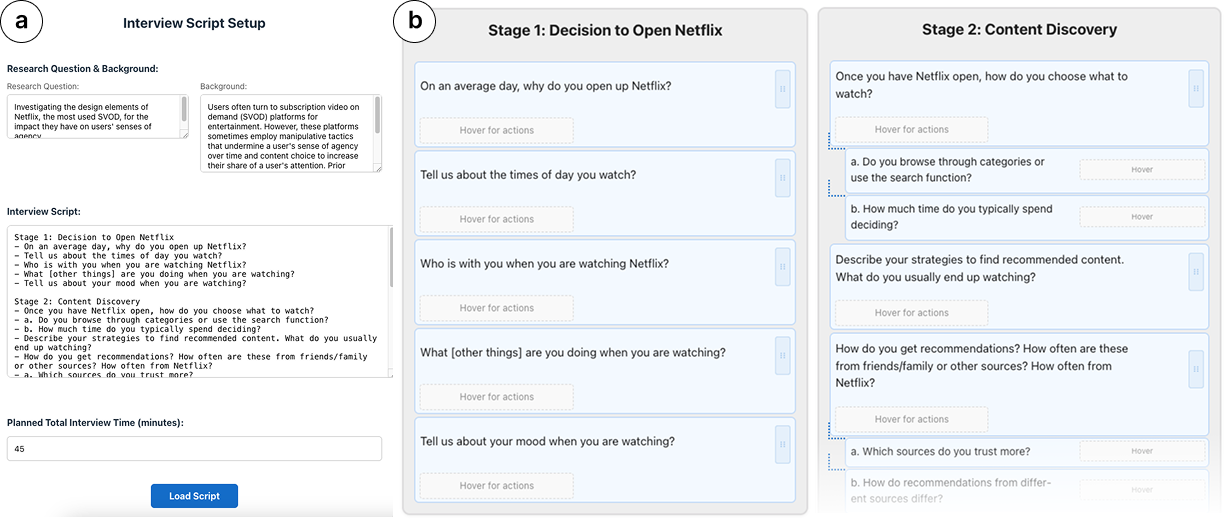}
    \caption{(a) Pre-interview setup page for uploading script and contextual information (research question, background introduction, planned interview time) (b) Enhanced interview script with visual hierarchy.}
    \Description{(a) Pre-interview setup page for uploading script and contextual information(research question, background introduction, planned interview time) (b) Enhanced interview script with visual hierarchy.}
    \label{fig:upload-script}
\end{figure*}

\section{Design Goals}
Based on the insights revealed from the formative study and design implications from previous literature ~\cite{liu2025envisioning}, we postulate that our interface should help interviewers manage interview flow while not being distracting and facilitate real-time data sensemaking, with the specific goals of: 

\subsection{[G1] Help Interviewers Manage the Interview Flow}

\paragraph{\textbf{[G1.1] }Enhancing interview script navigation with clear visual hierarchy}
Interviewers often struggled with script navigation due to limited cognitive bandwidth. As a result, they seldom adjust the script, worrying they might forget skipped questions or lose track of the order (F1). We plan to address this by presenting the script with a visual hierarchy that reflects the structure of the interview, highlights the ongoing question, and reveals which questions have not been covered by automatic conversation-question matching. 

\paragraph{\textbf{[G1.2]} Enhancing situational awareness with real-time time allocation visualizations}
It is hard for interviewers to mind their behavior during interviews (F3). The system should track interviewers' time allocation, such as time spent, topic coverage, and talking ratio, to enhance their situational awareness, avoiding inadvertently talking too much or spending excessive time on one topic, to ensure consistent data collection and the full expression of the interviewee's genuine thoughts.

\subsection{[G2] Facilitate Real-time Data Sensemaking }

\paragraph{\textbf{[G2.1]} Reducing effort for fine-grained information capturing }
As we observed in the formative study, non-expert interviewers have strong information-capturing needs, while manual note-taking may disrupt the natural conversation flow (F2, F4). Due to the nature of semi-structured interviews, the specific details the interviewee mentioned are more meaningful than coarse-based summaries of what has been talked about in general. Thus, precise recognition of user intent and extracting the details they want from the conversation is important. The system needs to enable interviewers to effortlessly express their intent for capturing information and deliver it in a concise manner for easy understanding and reviewing during the interview. 

\paragraph{\textbf{[G2.2]} Providing proactive reminders of noteworthy content}
Interviewers prefer a ``co-interviewer'' setting to alleviate the workload for making sense of the interviewee's responses, as well as having an additional perspective to ensure the comprehensiveness of the data collected (F4). We aim to leverage LLM agent as the ``co-interviewer'' to identify and highlight key points in the interview that merit attention or follow-up. The agent will be equipped with a background understanding of the research question and empirical knowledge of semi-structured interviews, to ensure the quality of the suggestions it makes. These suggestions are meant to alert interviewers to potentially overlooked details and expand their thoughts. Therefore, it is not feasible for users to initiate such requests. We want to have an ``always on'' assistant that listens to the interview conversation and presents their suggestions in a proactive way. Such suggestions should also be explainable to help interviewers make decisions on the next step.


\section{\system{}}

The following outlines \system{}’s three major components: the (1)
Interactive Script, (2) Visual Timer and (3) Mixed-initiative Information Capture. We then describe how a user can use \system{} for conducting a semi-structured interview.

\subsection{Interview Script with Visual Hierarchy (G1.1)}
\h{The system will turn the pure-text script into hierarchical UI components that reflects the structure of the interview (\autoref{fig:upload-script}). }This structure includes the stages, the main questions within each stage, and the subquestions that follow each main question. We derived it by reviewing eight recent papers (published within the past three years at HCI venues such as CHI and CSCW) that employed interviews as a research method and provided their scripts in appendices or supplementary materials (e.g.,~\cite{zhang2024sa}). There is no strict requirement on the format of the interview script; if the script does not have a clear structure, the system will automatically parse it into a structured representation. Moreover, users can directly drag to reorder questions within the panel.

After the interview begins, the system highlights the ongoing question in the script, fades completed questions to gray, and keeps the questions that have not yet been asked in their original light blue color. The system automatically monitors the conversation to detect the current question, while also allowing users to manually select a question to resolve potential detection errors.

\begin{figure}[!ht]
    \centering
    \begin{subfigure}[b]{0.52\linewidth}
        \includegraphics[width=\linewidth]{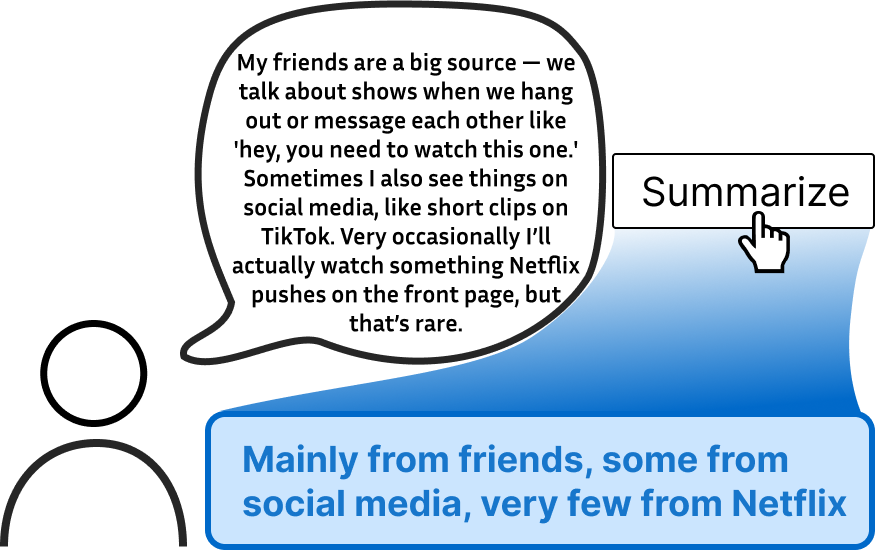}
        \caption{The system instantly generates a summary of the ongoing conversation when the user clicks the ‘Summarize’ button.}
        \Description{A highlighted “Summarize” button is pressed, and below it the system displays a short textual summary of the ongoing interview conversation, which the interviewer can review and edit as structured notes.}

        \label{fig:summary}
    \end{subfigure}
    \hfill
    \begin{subfigure}[b]{0.44\linewidth}
        \includegraphics[width=\linewidth]{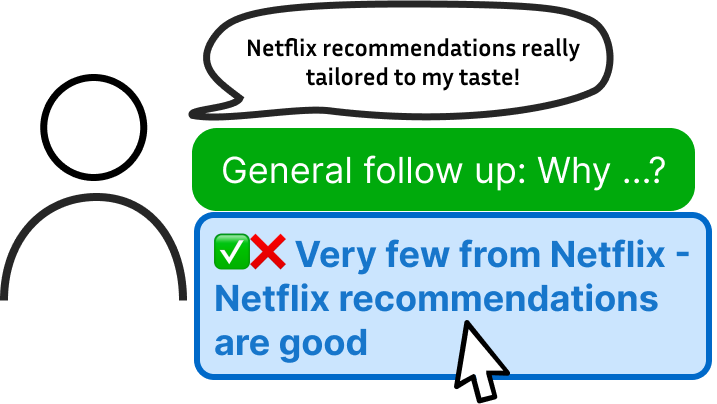}
        \caption{System suggests inconsistency with what the interviewee has said previously. User can hover on the tag to see further suggestion.}
        \Description{System suggests inconsistency with what the interviewee has said previously. User can hover on the tag to see further suggestion.}
        \label{fig:inconsistency}
    \end{subfigure}
    \caption{Mixed-initiative Information Capturing}
    \Description{Two screenshots showing the summarization and follow-up features.}
    \label{fig:features}
\end{figure}
\subsection{Visual Timer (G1.2)}
The visual timer(\autoref{fig:teaserFigure} b) makes the temporal dynamics of the interview visible. It visualizes time allocation across different topics as well as the speaking ratio between the interviewer and the interviewee. The timer also functions as a ``minimap'' of the complete interview script, providing an overview of the entire structure while indicating which stages have been completed and which remain.

\subsection{Mixed-initiative Information Capture (G2)}
\paragraph{``Manual Tagging'' Mode (G2.1)}
During the interview, the user can hover on the blank area under each question (\autoref{fig:teaserFigure} c) to click on ``create new tag'' to write down the content that does not come directly from the conversation, such as the interviewer's own thoughts, the non-verbal information like the interviewee's facial expression, etc.

\paragraph{``Click to Summarize'' Mode (G2.1)}
\h{The user can click on the ``Summarize'' button to let the system instantly generate an extractive summary of what the interviewee has said in the most recent dialogue (\autoref{fig:summary}).} The system will take the ongoing question into consideration when generating the summary, trying to extract the information related to the question at a fine-grained level. 
\begin{figure}[!ht]
    \centering
    \begin{subfigure}[t]{0.28\linewidth}
        \includegraphics[width=\linewidth]{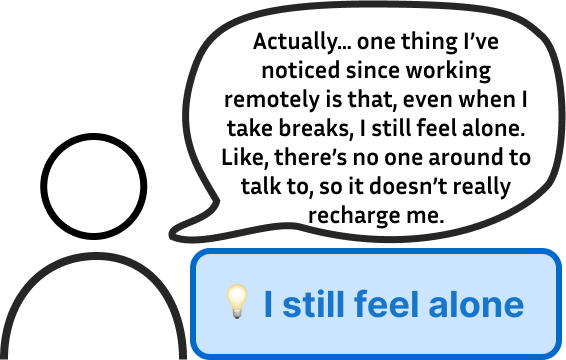}
        \caption{System suggests new relevant topic emerging from the conversation}
        \Description{ System suggests new relevant topic emerging from the conversation }
        \label{fig:new_theme}
    \end{subfigure}
    \hfill
    \begin{subfigure}[t]{0.32\linewidth}
        \includegraphics[width=\linewidth]{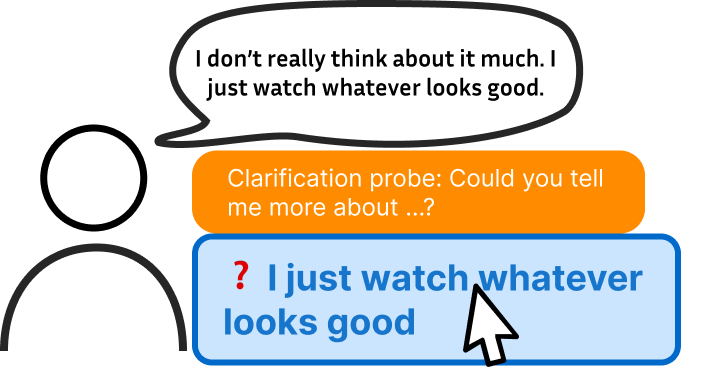}
        \caption{System suggests too general response from the interviewee, where a probe is suggested when hovering on it.}
        \Description{System suggests too general response from the interviewee, where a probe is suggested when hovering on it.}
        \label{fig:unclear}
    \end{subfigure}
    \hfill
    \begin{subfigure}[t]{0.32\linewidth}
        \includegraphics[width=\linewidth]{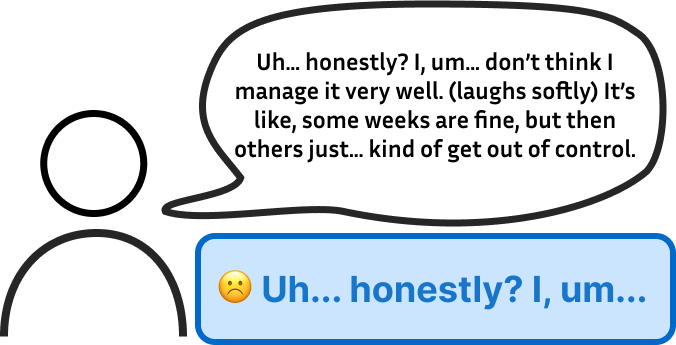}
        \caption{System suggests hesitation from the interviewee}
        \Description{System suggests hesitation from the interviewee}
        \label{fig:hesitates}
    \end{subfigure}
    \caption{Other Examples of the Co-interviewer Agent's Suggestions}
    \Description{Other Examples of the Co-interviewer Agent's Suggestions}
    \label{fig:interview_moments}
\end{figure}
\paragraph{``Co-interviewer'' Mode (G2.2)}
\label{sec:detecting-questions}
The system will proactively surface potential follow-up points (e.g., ``no specific steps described'' in  \autoref{fig:teaserFigure}c) that may be overlooked by the interviewer, drawing on the understanding of the conversation context, the broader interview goals, and the empirical knowledge about semi-structured interviews, which is based on literature on the structure of responsive interviews \cite{warren2002qualitative} We identified four common situations where interviewers often need to explore further. 
\label{scheme}
\h{\textit{Cases requiring probe: (1) when responses are vague or overly general, or rely on shared knowledge, unclear pronouns, gestures, or jargon (e.g., ``you know how it is'') (\autoref{fig:unclear}); (2) when the interviewee hesitates, self-corrects, or displays uncertainty, which may signal discomfort or deeper meaning (\autoref{fig:hesitates}). Cases requiring follow-up: (3) when new but relevant concepts or themes emerge beyond the immediate scope of the question (e.g., shifting from car seat use in a personal vehicle to problems on a school bus)(\autoref{fig:new_theme}); (4) when apparent contradictions or inconsistencies occur within the interviewee’s responses (e.g., ``ads don’t bother me'' vs. ``ads are so annoying'') (\autoref{fig:inconsistency})}. After coming up with the scheme, we invited two experts in qualitative research from our university to validate and complement the scheme by providing examples of each situation as few-shot demonstrations to LLM. }

The system self-evaluates the quality of generated suggestions and presents the ones of the best quality during natural pauses in the conversation. We design these suggestions to take an intermediate form, highlighting specific points in the conversation that warrant further exploration, along with guidance on how to follow up on them, without directly suggesting follow-up questions. This design choice aims to keep users cognitively engaged while reducing the risk of being misled by potentially inappropriate AI suggestions. Initially, only the detected situation is presented. If users want more actionable suggestions, they can hover over the current tag to see how it could be further explored (\autoref{fig:inconsistency} or \autoref{fig:unclear}). These suggestions are grounded in Hu et al.’s analysis of interviewers’ strategies for formulating follow-up questions~\cite{hu2024designing}, as well as prior literature on the structure of responsive interviews~\cite{warren2002qualitative}.

\subsection{Usage Scenario}
Here we present an imagined scenario about James, a student investigating users’ sense of agency with time and content choice on Netflix as his thesis project. The topic and script draw upon Schaffner et al.\cite{schaffner2023don}. James develops an interview script with main and follow-up questions grouped into stages and schedules a session with Amy, a Netflix user. Lacking experience, he decides to use \system{} for support.

\textit{Provide contextual information}
Before the interview, James inputs the research background, goal, complete script, and planned time. \system{} returns the script with a visual hierarchy (\autoref{fig:upload-script}), showing relationships between stages, main questions, and subquestions. For instance, under Stage 2: Content Discovery, the main question ``How do you get recommendations?'' is displayed with subquestions about sources such as friends, social media, or Netflix.

\textit{Navigate the script (G1.1)}
When the interview begins, James refers to the interactive script to stay oriented to the interview content. After asking ``Why do you use Netflix? On an average day, why do you open it?'', the system highlights the current question in yellow, while unanswered ones remain the original color and asked ones turn gray. This at-a-glance overview reassures him that he can skip ahead when Amy talks about recommendations and later circle back to cover missed questions.

\textit{Note-taking during interview (G2.1)}
When Amy gives a long answer about discovering shows, James hovers under the ongoing question and clicks the ``Summarize'' button. The system generates a focused summary: ``mainly from friends, some from social media, very few from Netflix.'' The conversation continues smoothly (\autoref{fig:summary}). Later, noticing Amy’s frown about autoplay, James creates a tag ``mixed feelings,'' reinforcing his impression of her ambivalence.

\textit{Identify and act on follow-up opportunities (G2.2)}
When Amy says recommendations are ``really tailored to my taste,'' \system{} detects a potential inconsistency with her earlier claim of rarely using them. It surfaces a compact tag (\autoref{fig:inconsistency}): ``very few from Netflix – Netflix recommendations are good.'' Hovering reveals more context, prompting James to ask, ``You feel the recommendations are good, but earlier you mentioned rarely using them?'' Amy then elaborates further.

\textit{Metacognition during interview (G1.2)}
Halfway through, James glances at the timer: half the time has been spent on Content Discovery while later stages remain untouched. He gently wraps up and transitions forward. The timer also shows James has spoken more than half the time. Realizing his closed-ended questions limit Amy’s elaboration, he reframes the next one to be more open-ended, stepping back to let her share more.

\section{Implementation}

\system{} is a web application developed with React\footnote{https://react.dev/} on the frontend and Firebase Functions\footnote{https://firebase.google.com/products/functions} on the backend. The following subsections describe the implementation of its core functions. The data flow of the system is shown in \autoref{fig:technical pipeline1}. All detailed prompts are included in Appendix A.2.

\h{
\paragraph{Retrieving Ongoing Question from the Script}
\label{dense-retriever}
Once the interview script is uploaded, GPT-4o parses its hierarchical structure, including stages, main questions, and subquestions. Audio of the ongoing conversation is streamed via WebSocket\footnote{https://developer.mozilla.org/en-US/docs/Web/API/WebSockets\_API}. The question retrieval module then processes both inputs. The audio data first goes through AssemblyAI’s Speech-to-Text API\footnote{https://www.assemblyai.com/docs/speech-to-text/universal-streaming} to be transcribed in real time. For every 50 words, we use SpaCy\footnote{https://spacy.io/} to perform sentence boundary detection to avoid cutting sentences mid-utterance. The 50-word window was chosen through pilot testing to balance latency and accuracy. We use a dense retriever to retrieve the question from the script that is the most relevant to the ongoing conversation. Specifically, we encode the dialogue and the interview questions into text embeddings using OpenAI’s \texttt{text-embedding-3-small} model and compute the similarity between them, selecting the question whose embedding has the highest cosine similarity to the ongoing conversation with a threshold of 0.5. To provide explainability in the system's detection results, the similarity score $s \in [0,1]$ is mapped to the opacity of the color for highlighting the ongoing question ($\text{opacity} = s^{2}$), where low confidence produces a near-white background, and higher confidence saturates toward yellow. 
}

\begin{figure*}[!ht]
    \centering
    \includegraphics[width=0.85\linewidth]{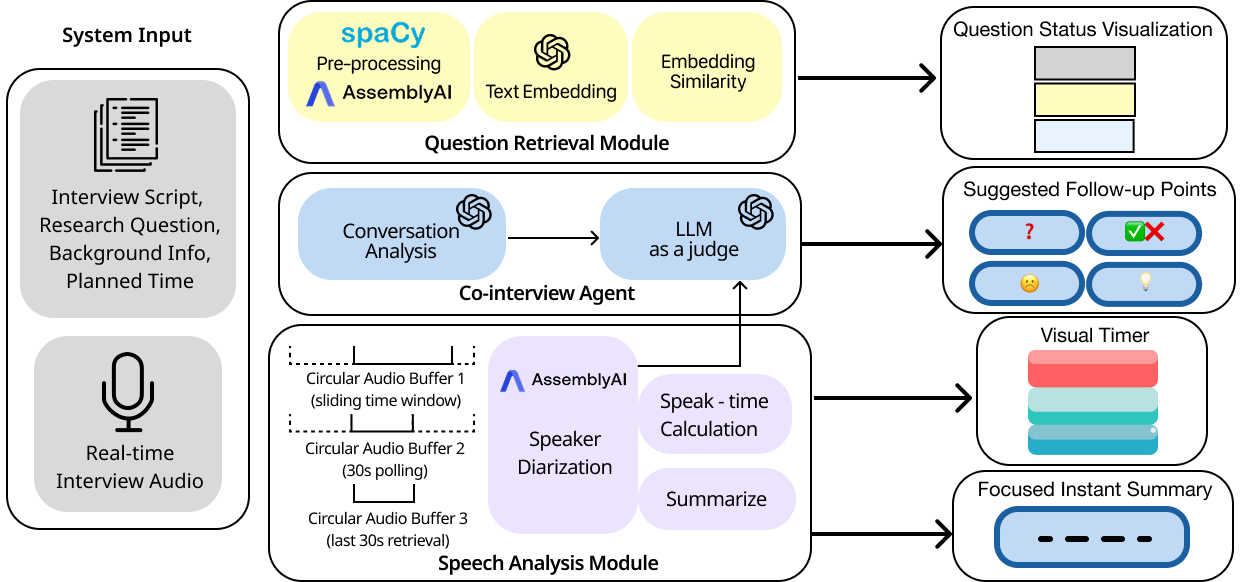}
    \caption{System Data Flow}
    \Description{Diagram of the InterFlow system pipeline for AI-assisted semi-structured interviews. On the left, the system takes two inputs: an interview script containing research questions, background information, and planned time, and the real-time interview audio. The script is processed by a question retrieval module using spaCy for preprocessing and text embeddings, which drives a question status visualization and generates suggested follow-up points for the interviewer. In parallel, the audio is handled by a speech analysis module with circular audio buffers, speaker diarization, speaking-time calculation, and on-the-fly summarization of recent conversation. A co-interview agent combines conversation analysis with a large language model acting as a judge to evaluate the dialogue and surface points worth probing. The outputs include a visual timer showing interview progress and speaking balance, and a focused instant summary that presents concise, up-to-date notes to support the interviewer during the session.}
    \label{fig:technical pipeline1}
\end{figure*}

\h{
\paragraph{Generating Instant Summary with User-Specified Focus}}
The frontend implements audio capture using WebRTC\footnote{https://webrtc.org/} and the Web Audio API\footnote{https://developer.mozilla.org/en-US/docs/Web/API/Web\_Audio\_API}. We use the circular audio buffer 3 for storing the most recent conversation. When a request is triggered, the circular buffer supplies the last 30 seconds of audio. This piece of audio data is processed through AssemblyAI’s LeMUR framework~\cite{assemblyai_lemur2023} and the \texttt{claude-3.5-sonnet} model to generate a customized extractive summary of the interviewee’s response. We chose Claude-3.5-sonnet for its advantage of short response time. In our implementation, the end-to-end latency from when a user clicks to mark a segment until the summary result is presented is approximately 4–5 seconds.

\paragraph{Co-interview Agent}
The Co-interview Agent is built on OpenAI’s real-time API\footnote{https://platform.openai.com/docs/guides/realtime-models-prompting}, which provides embedded Voice Activity Detection (VAD). We configured VAD in semantic mode to ensure segmentation decisions rely not only on acoustic pauses but also on semantic cues, reducing the likelihood of cutting off an unfinished utterance and enabling the model to respond at natural breakpoints. \h{We provide the scheme for typical situations that require further exploration along with few-shot demonstrations (see \autoref{scheme}  for details) to the \texttt{gpt-realtime} model to let it extract relevant content in the interview conversation that falls into the scheme. To calibrate suggestion frequency and quality, we used the ``LLM-as-a-judge'' framework~\cite{gu2024survey} to evaluate accumulated results over sliding time windows and select the best one to be presented to users. }Let $t_i$ denote the arrival time of the $i$-th tag; the circular audio buffer $\mathcal{B} = \{s_j, s_{j+1}, \dots, s_k\}$ is constructed such that
\[
t_{m+1} - t_m < 10 s \quad \text{for all } j \leq m < k.
\]
The buffer extends as long as results arrive within 10 seconds of each other; once a gap exceeds this interval, the buffer is closed and sent, together with the transcript in the window, to GPT-4o for evaluation (see \autoref{fig:technical pipeline1}). Outputs are rated on correctness, specificity, and salience, and the highest-rated result is surfaced to the user when a conversational pause is detected. 

\paragraph{Nearly Real-time Talking Ratio Visualization}
A separate circular buffer is maintained and updated every 30 seconds with the latest audio data. The buffer is processed using AssemblyAI’s API\footnote{https://www.assemblyai.com/docs/guides/talk-listen-ratio} for speaker diarization and talk-time calculation. The accumulated speaking time of the interviewer and interviewee within the current stage is then computed, and the ratio is updated on the progress bar to reflect conversational balance.

\paragraph{Coordination Between User Preference and Agent Behavior}
To prioritize explicit user input, a pause mechanism is integrated into question detection. When the interviewer clicks on a question to mark it as ongoing, the automatic detection module is suspended for 15 seconds. If no further manual input occurs during this period, the system resumes real-time transcription and question retrieval. Additionally, when notes are recorded manually or a summary of recent dialogue is requested, the resulting content is passed to the Co-interview Agent as part of its context. Incorporating user-provided input in this way allows the agent to refine its evaluations and avoid redundant or repetitive suggestions.

\section{Evaluation}

\h{
We conducted a lab study with twelve interviewer-interviewee pairs to evaluate \system{} against a baseline system (\autoref{fig:baseline system}). Our evaluation aimed to address three research questions:

RQ1: To what extent does \system{} support interviewers in managing interview flow?

    RQ2: To what extent does \system{} support interviewers’ real-time data sensemaking?

RQ3: What concerns or potential risks arise from using \system{} in interview settings?
}

\subsection{Participants}

\h{
\subsubsection{Interviewer}
For the recruitment of interviewers, we distributed a call for participation through university email lists and social media. We asked the prospective participants to rate their interview expertise level and collected the frequencies they had previously conducted interviews to obtain a reliable assessment, which is the same as formative study. 
We screened all prospective interviewers and excluded one who reported expert-level proficiency (rated 7 in the expertise level and had interview experience more than 20 times). The average self-rated expertise of final participants is 3.55 (SD=0.72), with 4/12 had interviews for 0-5 times, 2/12 had 5-10, 1/12 had 10-15, 2/12 had 15-20, 3/12 had over 20 times. 
The detailed demographic information about the final twelve interviewers are presented in \autoref{tab:evaluation-demographics} in Appendix.
\subsubsection{Interviewee}
Twelve interviewees were recruited through personal networks and snowball sampling, and matched to interviewers by selecting individuals who could meaningfully discuss the topics chosen by the interviewers. The demographic information about interviewee participants is presented in \autoref{tab:evaluation-demographics-interviewee} in Appendix.
}

Of the twelve study sessions, eight were conducted remotely via Zoom and four were conducted in person. All the interviewee participants were aware of the interviewer's usage of AI. Each interviewer participated for approximately 100 minutes and received \$40 or an equivalent Amazon gift card. Interviewees participated for about one hour and received \$20 gift card or its equivalent, reflecting their shorter time commitment and less demanding role.\footnote{The user study was approved by the Institutional Review Board (IRB) at the local university.}

\begin{figure*}[!ht]
    \centering
    \includegraphics[width=\linewidth]{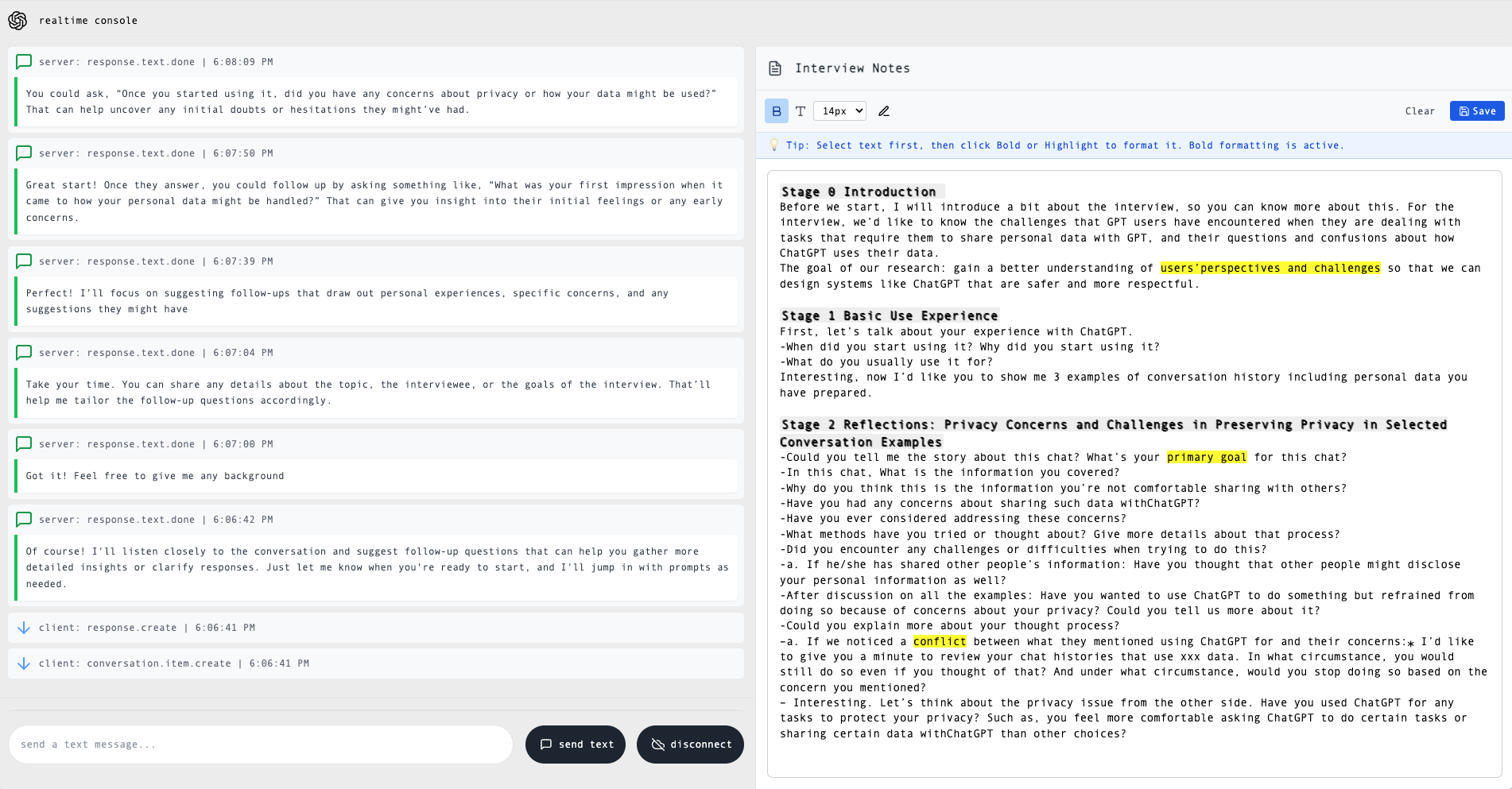}
    \caption{Baseline system contains a chat panel (left) powered by OpenAI Realtime API and a text editor (right). Before the interview starts, users can provide prompts to configure the AI assistant. The AI assistant will listen to the conversation and proactively provide help in text format during the interview.}
    \Description{Baseline system contains a chat panel (left) powered by OpenAI Realtime API and a text editor (right). Before the interview starts, users can provide prompts to configure the AI assistant. The AI assistant will listen to the conversation and proactively provide help in text format during the interview.}
    \label{fig:baseline system}
\end{figure*}

\h{
\subsection{Baseline}
The design of baseline (\autoref{fig:baseline system}) system is inspired by interviewers' common practices of presenting the interview script and taking notes using a text editor. 
To ensure a fair comparison with \system{} and test whether the benefits or shortcomings of \system{} arise from our special design or the general capabilities of AI, we integrate the same model (Open AI's gpt-realtime) into the baseline system. 
However, instead of the specially designed co-interview agent, we let users prompt the model themselves before starting the interview to provide contextual information and express their preferences for follow-up suggestions. The other core features like interactive script with visual hierarchy, the instant summary with user-specified focus, and the visual timer, were not provided. Users could manually adjust the visual hierarchy and take notes using the text editor, and they can view the time on screen in the baseline system. 

}

\subsection{Materials}
\label{efforts for enhancing realism}
\h{We prepared three interview topics of comparable length and difficulty, trying to enable interviewers to choose a topic they are interested in and had some familiarity with, to better approximate a realistic interview setting. To keep the study duration manageable—given that each participant will conduct two interviews—we manually shortened the scripts to reduce participant fatigue. The topics were: (1) privacy concerns in using large language models (10 questions, three stages, adapted from Zhang et. al's work \cite{zhang2024sa}), (2) sense of agency in time and content choice on Netflix (10 questions, three stages, adapted from Schaffner et. al's work \cite{schaffner2023don}), and (3) usage patterns and behaviors in video conferencing for presentations (14 questions, five stages, adapted from a script provided by an industry UX designer). The detailed information is included in Supplementary Materials.}


To help participants prepare, we sent them the selected interview scripts at least one day before the study and encouraged them to make adjustments to suit their preferences. In practice, eight participants selected topic (1) (four times in baseline, four times in \system{}), five selected topic (2) (two times in baseline and three times in \system{}), and eleven selected topic (3) (six times in baseline and five times in \system{}). We counterbalanced the assignment of topics across the two system conditions so that each topic appeared in both the baseline and \system{} conditions (almost) equally. 

\subsection{Procedure}

A single session was divided into three main stages, described as follows.

\subsubsection{Pre-study survey (\textasciitilde5 min)}
At the beginning of the session, participants were briefed on the purpose of the study and asked to read and sign a consent form. After obtaining consent, they completed a brief survey to provide background information regarding their familiarity and interest with the two interview topics they selected.

\subsubsection{Main study (\textasciitilde75 min)}
Participants engaged in two separate interview sessions, one with \system{} and one with the baseline system. We adopted a within-subject design and counterbalanced the order of conditions across participants, with six pairs starting from the baseline and six pairs starting from \system{}. Each session began with a 5-minute tutorial, during which participants were introduced to the key features of the current system and were given a brief hands-on trial to familiarize themselves with the interface. The tutorial was followed by a 25-minute semi-structured interview task, where the interviewer used the assigned system to conduct the interview with the paired interviewee. The sessions were designed to be independent of each other, and the same interviewee was paired with the interviewer in both sessions to control for interpersonal variation. After each session, both the interviewer and the interviewee completed a post-session survey. The participants had the option of a 10-minute break between the sessions.

\subsubsection{Exit interview (\textasciitilde20 min)}
Lastly, we conducted a 20-minute semi-strucutred interview with the interviewee to ask about their feelings of interviewer's usage of external tools, and a 15-minute one with the interviewer to ask about the difference between their experience in the two conditions, (1) the differences between their experiences in the two conditions, (2) their perceptions of the systems’ usefulness and usability, and (3) their concerns or reservations about using such tools in real interview settings.

\subsection{Measurements}
\h{To assess the usability and usefulness of \system{}, we drew on multiple data sources, including interaction logs, session duration, the number of questions asked, quantitative results on the system performance, plus post-session surveys and interviews from participants.}

We recorded participants’ interactions with \system{}, together with the content returned from the system. From these logs, we quantified how often interviewers manually took notes, how often they clicked to request AI for summarizing the conversation, how many proactive suggestions are generated, whether and how they utilized the suggestions. \h{We measured the system's latency and accuracy in detecting the ongoing question, the latency of returning instant summaries, and the quality of suggestions for further exploration.
}

After each session, both interviewers and interviewees completed a post-session survey. For interviewers, the survey examined the following dimensions: (1) Cognitive load: measured using the NASA Task Load Index (NASA-TLX)~\cite{NASA}.
(2) System usability: measured using the System Usability Scale (SUS)~\cite{brooke1996sus}.
(3) Managing interview flow: how well the system enhanced situational awareness and supported navigation of the interview script.
(4) Real-time data sensemaking: how well the system facilitated capturing key information and surfacing points for elaboration.
(5) Feature-specific assessments: perceptions of visualization of interview structure and status, visual timer, AI-assisted note-taking, and system-suggested follow-ups.
(6) Concerns and risks: potential issues identified in prior literature, including distraction from the conversation, hindering the development of interview skills, affecting data authenticity, and impacting interviewee experience.

For interviewees, the survey focused on their experience of being interviewed, including the perceived smoothness of the interview and their comfort with the interviewer’s use of the system. The detailed survey questions are included in Appendix.

\begin{table*}[]
\begin{tabular}{lcccccccc}
\hline
\multicolumn{1}{c}{\textbf{No.}} & \multicolumn{2}{c}{\textbf{Q-Detect}} & \multicolumn{2}{c}{\textbf{Summary}} & \multicolumn{2}{c}{\textbf{\% Noisy Suggestions}} & \multicolumn{2}{c}{\textbf{Quality of Suggestions}} \\ \cline{2-9} 
                                 & \textbf{Acc.}     & \textbf{Lat.}     & \textbf{Acc.}     & \textbf{Lat.}    & \textbf{\system{}}  & \textbf{Baseline} & \textbf{\system{}} & \textbf{Baseline} \\ \hline
P1 (remote)                      & 0.67              & 8.1s              & 0.55              & 2.9s             & 40\%                             & 5\%               & 3.00                            & 1.85              \\
P2 (in person)                   & 0.61              & 10.5s             & 1.00              & 3.9s             & 29\%                             & 35\%              & 2.42                            & 2.30              \\
P3 (remote)                      & 0.60              & 9.6s              & 0.65              & 3.6s             & 33\%                             & 0\%               & 2.75                            & 2.60              \\
P4 (in person)                   & 0.64              & 11.1s             & 0.80              & 5.1s             & 36\%                             & 45\%              & 2.11                            & 2.90              \\
P5 (remote)                      & 0.50              & 7.8s              & 1.00              & 6.0s             & 0\%                              & 10\%              & 3.00                            & 3.00              \\
P6 (remote)                      & 0.67              & 6.9s              & 0.55              & 3.5s             & 0\%                              & 25\%              & 2.00                            & 2.40              \\
P7 (remote)                      & 0.57              & 11.3s             & 0.60              & 4.7s             & 27\%                             & 50\%              & 2.75                            & 2.80              \\
P8 (remote)                      & 0.54              & 5.4s              & 1.00              & 4.9s             & 20\%                             & 15\%              & 2.17                            & 3.00              \\
P9 (in person)                   & 0.58              & 7.7s              & 0.54              & 3.6s             & 38\%                             & 30\%              & 1.20                            & 2.50              \\
P10 (in person)                  & 0.60              & 9.3s              & 1.00              & 4.3s             & 50\%                             & 20\%              & 1.57                            & 2.70              \\
P11 (remote)                     & 0.56              & 12.4s             & 0.70              & 5.3s             & 40\%                             & 40\%              & 1.80                            & 3.00              \\
P12 (remote)                     & 0.50              & 9.5s              & 0.88              & 3.0s             & 33\%                             & 53\%              & 1.71                            & 2.72              \\ \hline
\textbf{Avg}                     & 0.58              & 8.9s              & 0.77              & 4.2s             & 29\%                             & 27\%              & 2.21                            & 2.66              \\ \hline
\end{tabular}
  \caption{Technical performance of \system{}. Acc. means accuracy and lat. means latency. \% Noisy suggestions indicates the percentage of suggestions that contain factual errors. The quality of suggestions is rated by experts on a 1–3 scale. 1 indicates low quality, and 3 indicates high quality. The inter-rater reliability (IRR) between the two experts was 0.64 for \system{}, indicating substantial agreement. For the baseline, the IRR was 0.54, reflecting moderate agreement. }
  \label{tab:system-performance}
\end{table*}

\subsection{Analysis}
We collected the study recordings, the interaction logs, and survey responses to perform quantitative and qualitative analysis. We conducted a thematic analysis ~\cite{terry2017thematic} on the transcripts of the exit interviews. Two authors independently coded the data to identify recurring patterns related to participants’ perceptions with system features and their reasoning behind interview decisions. They then met to resolve conflicts and update codes. These codes were then iteratively organized into more coherent, higher-level themes through discussions within the whole research team. 

Additionally, we annotated the recording of each study session, labeling system events, user requests and the system’s responses. We evaluated \system{}'s technical performance on its accuracy and latency of detecting ongoing question and instant summary, and the quality of suggestions for further exploration of both baseline system and \system{}:

\paragraph{Accuracy and latency of detecting the ongoing question}
    The accuracy is calculated as
\begin{equation*}
    \text{Question Detection Accuracy} =
\frac{\#\ \text{correctly detected questions}}
{\substack{\#\ \text{questions in the script} \\
\text{that were asked by the interviewer}}}
\end{equation*}

    We treat all questions whose status was manually indicated by the user as system detection failures, because such instances imply that the system either responded too slowly or detected an incorrect question. Latency is defined as the time interval between the moment the interviewer finishes asking a question and the moment the system highlights the corresponding question. To compute accuracy and latency, the third author reviewed the study recordings of \system{}, annotated each instance where the interviewer asked a scripted question, determined whether the system identified it correctly, and measured the detection latency for all correctly detected questions. When encountering uncertain instances, the third author discussed with the first author to determine the result.

\paragraph{Accuracy and latency of returning instant summaries with user-specified focus}
    
    The accuracy is defined as the proportion of system-generated summaries that successfully captured what the interviewee said surrounding the moment when interviewer requested the summary. Latency is defined as the time interval between the moment the interviewer clicked on \texttt{Summarize} and the moment the system returned the summary. To derive these measures, the third author located each summary request in the study recordings, and examined the surrounding dialogue to give boolean label of whether it captured the key information, and recorded the time the system took to produce the summary. The first author verified the results by reviewing the annotations in the video.

\paragraph{Quality of suggestions for further exploration}

We recruited two experts in qualitative research from local university to evaluate the factual accuracy and quality of the follow-up suggestions of \system{} and baseline. One expert is a postdoctoral researcher in HCI, and the other is a postdoctoral researcher in Public Health. Each suggestion was rated on a 0–3 scale, where 0 indicates that the suggestion contained a factual error. Suggestions without factual errors were rated from 1 to 3 based on its quality, defined by whether the experts considered them meaningful for the interviewer at the moment when it appeared. We also asked the experts to explain the reasons behind any low ratings they assigned. We let the two experts independently review the study recordings and rate on the suggestions of both \system{} and the baseline, with condition information removed. Non-agreement cases on factual error were discussed to resolve conflicts. We then calculate the inter-rater reliability on the ratings of non-noisy suggestions (the ones do not contain factual error) using Cohen's Kappa with quadratic weight ~\cite{GISEV2013330}, and calculated the average rating to reflect the overall quality of suggestions.


\section{Results}
We conducted a Mann–Whitney test on interviewers’ familiarity with the script and their interest in the topics. Their familiarity with the script was moderate in both conditions (baseline: M = 4.67, SD = 1.65; \system{}: M = 4.33, SD = 1.29), while their interest in the topics was relatively high (baseline: M = 5.19, SD = 2.24; \system{}: M = 6.08, SD = 1.71). The test revealed no significant differences across topics. The results indicate that our study is relatively close to realistic scenario and there is no confounding effect introduced by the choice of topics. 


\begin{figure*}[!ht]
    \centering
    \includegraphics[width=\linewidth]{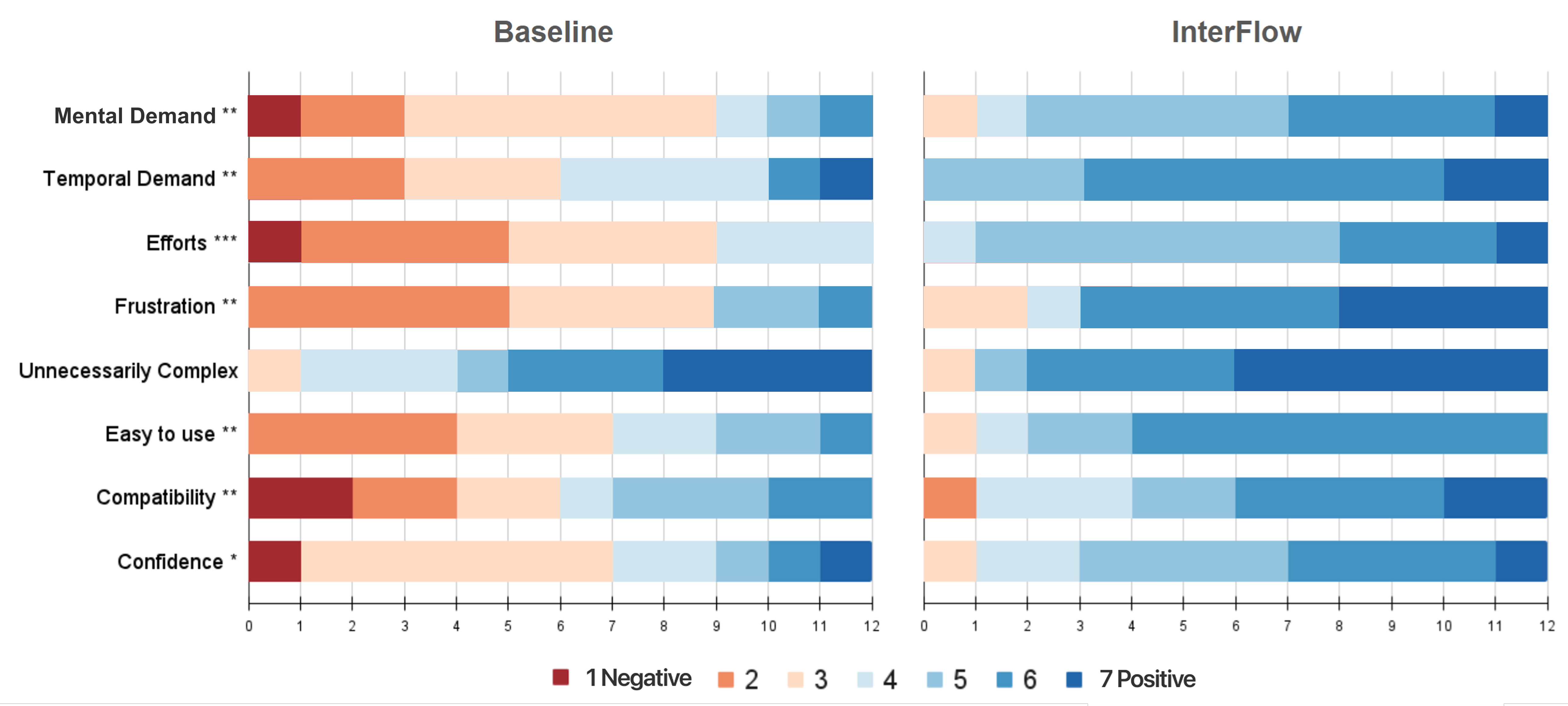}
    \caption{Distribution of user ratings for System Usability and Cognitive Load ( $p < 0.05$ is marked with *, $p < 0.01$ is marked with **, and $p<0.001$ is marked with ***)}
    \Description{.}
    \label{fig:system usability}
\end{figure*}

\subsection{Quantitative Results on System Performance}

\autoref{tab:system-performance} shows the technical performance of the core features of \system{}, including highlighting ongoing questions, instant summaries, and suggestions for further exploration.

The average accuracy of question detection is 58\%, and the average latency is 8.9 seconds. The question-detection errors consisted of two primary types: cases in which the system highlighted an incorrect question (around 45\%) and cases in which the system failed to respond before the user manually clicked the question to indicate its status (around 55\%). These errors typically occurred when interviewers rephrased or adapted the wording of the scripted questions, which reflects that the current dense retrieval method we adopted (see \autoref{dense-retriever} for details) might not be robust enough in dynamic interview settings.

The average accuracy for the summary function is 77\%, and the average latency is 4.2 seconds. Summary failures occurred when audio quality degraded, which sometimes occurred in remote sessions. \system{} would miss portions of the utterance; for example, when the interviewee mentioned using AI to improve writing and generate visuals, the system only captured the generation of visuals. Moreover, when only one person spoke at a time, \system{} struggled to reliably perform speaker diarization (e.g., the system returned ``insufficient context'') and therefore failed to provide a summary of what the interviewee said.

The percentage of noisy suggestions for \system{} and Baseline is 29\% and 27\%, respectively. One reason for noisy suggestions, shared by both systems, is the inherent hallucination problem of LLMs. 
For example, when an interviewee described their past experience of feeling sick, the system inferred this as hesitation or discomfort with the interview process—an interpretation that was inconsistent with the actual situation—and generated a follow-up suggestion based on this incorrect inference. The second type of error is incorrectly categorizing the conversational cue, which is particular with \system{}. For instance, when an interviewee mentioned wanting to create ``visually appealing slides,'' the system labeled this as a new theme, while such a statement is closer to an overly general situation and better suited for prompting more specific elaboration rather than opening a new thematic branch. 

The quality ratings of suggestions for \system{} and Baseline are 2.21 and 2.66, respectively. Both systems produced suggestions of moderate to high overall quality. Suggestions that received low expert ratings in \system{} were either less relevant to the interview goal or did not contain a substantive opportunity for further exploration. One example would be the interviewee mentioned they would save Netflix recommendations into the watch later list, and the system extracts ``watch later list'' as the excerpt that relates to a new theme. ``It doesn't seem to fall into the current focus of content discovery, but the interviewer might ask more about it later when they have time, explained one expert.

Low-rated suggestions from the baseline system were typically irrelevant to the ongoing conversation, overly leading, or resting on strong assumptions. For instance, when an interviewee described how declaring personal information to LLM helped the model provide more personalized responses that made things easier, the baseline system prompted: \textit{``Do you think knowing more about how your data is stored or used would change how you interact with these tools? Or do you feel like the benefits still outweigh those concerns for you?''} Experts found this problematic because the question was overly directive-``it seems to be engineered to get a certain response from the interviewee, and it implicitly assumed that the interviewee had not looked into how their data was managed, an assumption that might be incorrect.''

It is worth noting that \system{} exhibited a relatively higher proportion of noise and received lower quality ratings for its suggestions. We attribute this to the requirement of generating suggestions grounded in our predefined scheme. To do so, the model must accurately interpret the ongoing interview and decide if it is relevant to the situations in the scheme. This places higher demands on its instruction-following and reasoning capabilities. In contrast, the baseline system produces more generic follow-up suggestions, making it less susceptible to errors arising from misclassification or misalignment between the conversation context and the intended situation category. 
We will further discuss the discrepancies between user perceptions and expert evaluations of suggestions in \autoref{qualitative results of follow-up suggestions}.

\subsection{System Usability and Cognitive Load}
The user ratings for system usability and cognitive load are shown in \autoref{fig:system usability}. \h{During the onboarding process, we observed a low learning curve of the baseline system. All users managed to use the system after the 5- minute tutorial. While for \system{}, several participants were still unsure about how to trigger certain features in \system{}. However, after the two interview sessions, they rated \system{} significantly easier to use (Baseline: M = 3.42, SD = 1.38; \system{} M = 5.42, SD = 1.00; \textit{p} = .004, \textit{d} = -1.25) and more functionally integrated (Baseline: M = 3.58, SD = 1.83; \system{} M = 5.17, SD = 1.47; \textit{p} = .002, \textit{d} = -1.28). Despite offering more features, \system{} was considered straightforward and cohesive.}

Cognitive load ratings were also significantly lower with \system{}: mental demand (Baseline: M = 4.83, SD = 1.34; \system{} M = 2.75, SD = 1.06; \textit{p} = .002, \textit{d} = 1.44), temporal demand (Baseline: M = 4.33, SD = 1.56; \system{} M = 2.08, SD = 0.67; \textit{p} = .004, \textit{d} = 1.28), effort (Baseline: M = 5.25, SD = 0.97; \system{} M = 2.67, SD = 0.78; \textit{p} < .001, \textit{d} = 1.87), and frustration (Baseline: M = 4.83, SD = 1.40; \system{} M = 2.33, SD = 1.50; \textit{p} = .002, \textit{d} = 1.24). Participants attributed this to fine-grained summarization that reduced manual typing and a visual script hierarchy that streamlined tracking. As P10 explained: ``With \system{}, I could focus on listening while the system handled tracking and note-taking, which made the process much less stressful.''

\begin{figure*}[t]
    \centering
    \includegraphics[width=0.75\linewidth]{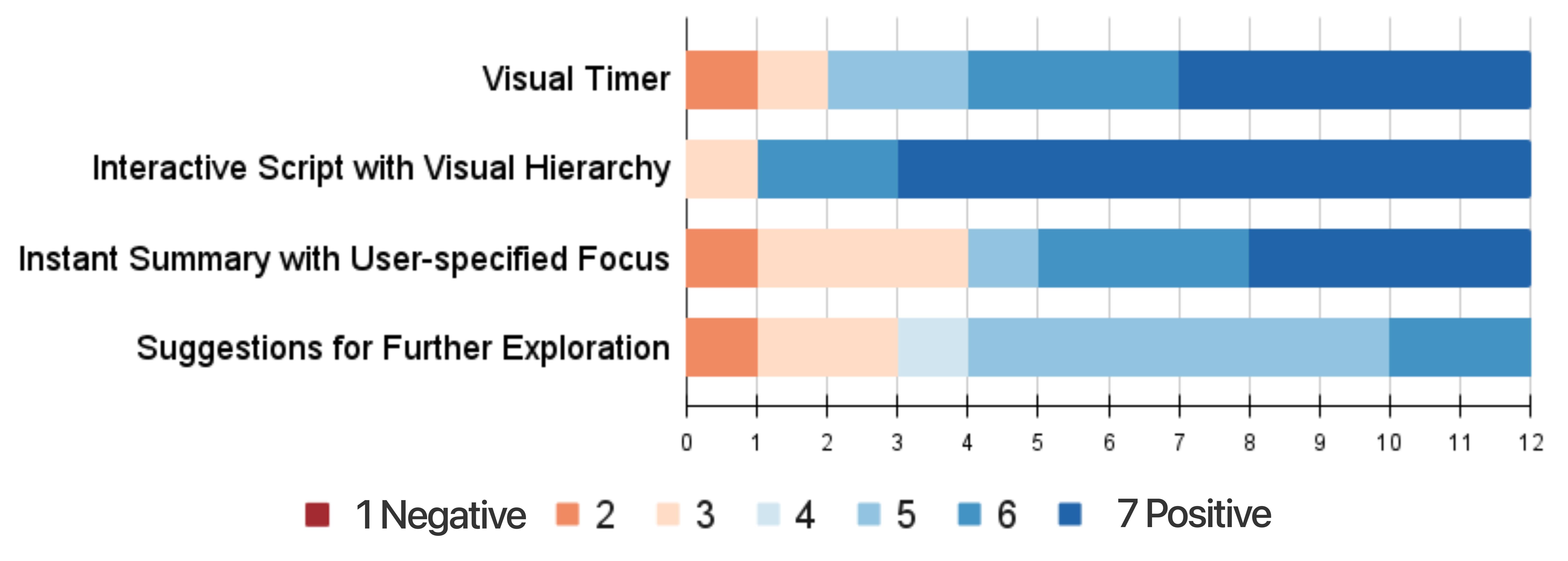}
    \caption{Distribution of user ratings on the degree of satisfaction for each feature of \system{}.}
    \Description{Distribution of user ratings on the degree of satisfaction for each feature of \system{}.}
    \label{fig:rating-features}
\end{figure*}

\begin{figure*}[!ht]
    \centering
    \includegraphics[width=\linewidth]{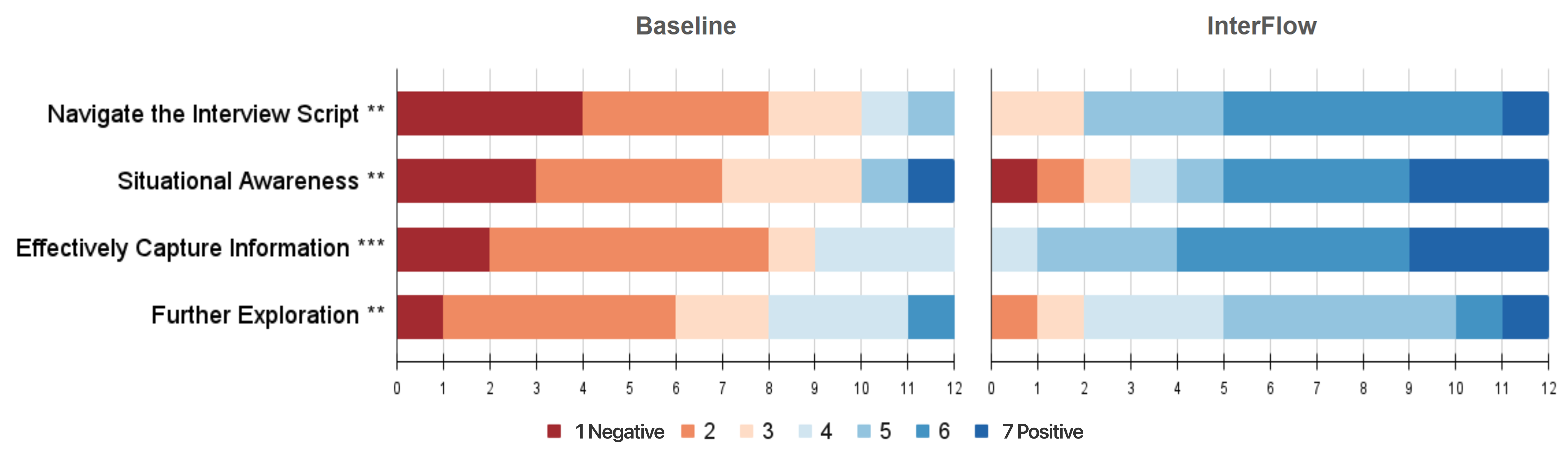}
    \caption{Distribution of user ratings for Managing Interview Flow and Realtime Data Sensemaking ( $p < 0.01$ is marked with ** and $p<0.001$ is marked with ***)}
    \Description{Distribution of user ratings for Managing Interview Flow and Realtime Data Sensemaking ( $p < 0.01$ is marked with ** and $p<0.001$ is marked with ***)}
    \label{system usefulness}
\end{figure*}

\subsection{How well does \system{} support interviewers in managing the interview flow?}
\subsubsection{Navigate the Interview Script}
Participants reported that \system{} significantly improved script navigation (Baseline: M = 2.25, SD = 1.29; InterFlow: M = 5.33, SD = 1.23; \textit{p} = .002, \textit{d} = -1.49). Although the system automatically detected the ongoing question and updated the visual hierarchy, interviewers often highlighted questions manually (41.4\% on average), usually as a prompt before asking. This reflected the system’s need to process sufficient dialogue before detecting the active question. Despite this, the feature was rated highly (M = 6.50, SD = 1.67), see \autoref{fig:rating-features}.

Participants expressed mixed views on manual highlighting. Some valued it as an orienting aid—``Such a small act prevents me from getting lost in a script full of text'' (P9)—while others preferred automation and found manual clicking burdensome when highlights lagged. The feature was particularly helpful for managing follow-up questions, as the hierarchy clearly displayed the relationship between main and follow-up questions based on that, supporting on-the-spot decision-making. By contrast, in the baseline condition, participants tracked progress with a text-editor cursor, which offered little visual salience and required more effort.

\subsubsection{Situational Awareness}
\h{The interviews conducted using \system{} lasted 20.33 minutes on average (SD=4.07), and the interviews conducted using baseline system lasted 17.58 (SD=5.85) minutes on average, with no statistically significant difference according to a paired t-test(\textit{p} = .08). In both conditions, the same participant exceeded the planned interview duration: 2 minutes with InterFlow and 9 minutes with the baseline system.}
Participants reported that \system{} significantly enhanced their situational awareness (Baseline: M = 2.67, SD = 1.78; \system{} M = 5.00, SD = 2.04; \textit{p} = .009, \textit{d} = -1.07). The visual timer was rated positively (M = 5.67, SD = 1.67), see \autoref{fig:rating-features}. We observed participants often glancing at it during natural pauses—for example, after completing a stage or following a long response. At these points, the timer served as a checkpoint for assessing elapsed time in a stage and remaining time overall. This awareness frequently shaped behavior: some interviewers wrapped up lengthy topics, while others slowed down to avoid rushing. As P2 explained, ``It helped me balance topic coverage and conversational depth. I tend to spend a lot of time on a certain topic when the interviewee is talkative about it.'' The timer thus acted not only as a reminder but also as a scaffold for metacognition, supporting pacing and attention distribution.

The speaking-ratio visualization, embedded in the time allocation display, provided an overview of interviewer vs. interviewee's speech balance. However, some participants found it too coarse: it updated infrequently and only at the stage level, limiting real-time usefulness. P6 suggested finer-grained sequential views, such as turn-taking points and speaking-turn lengths.

\begin{figure*}[!ht]
    \centering
    \includegraphics[width=\linewidth]{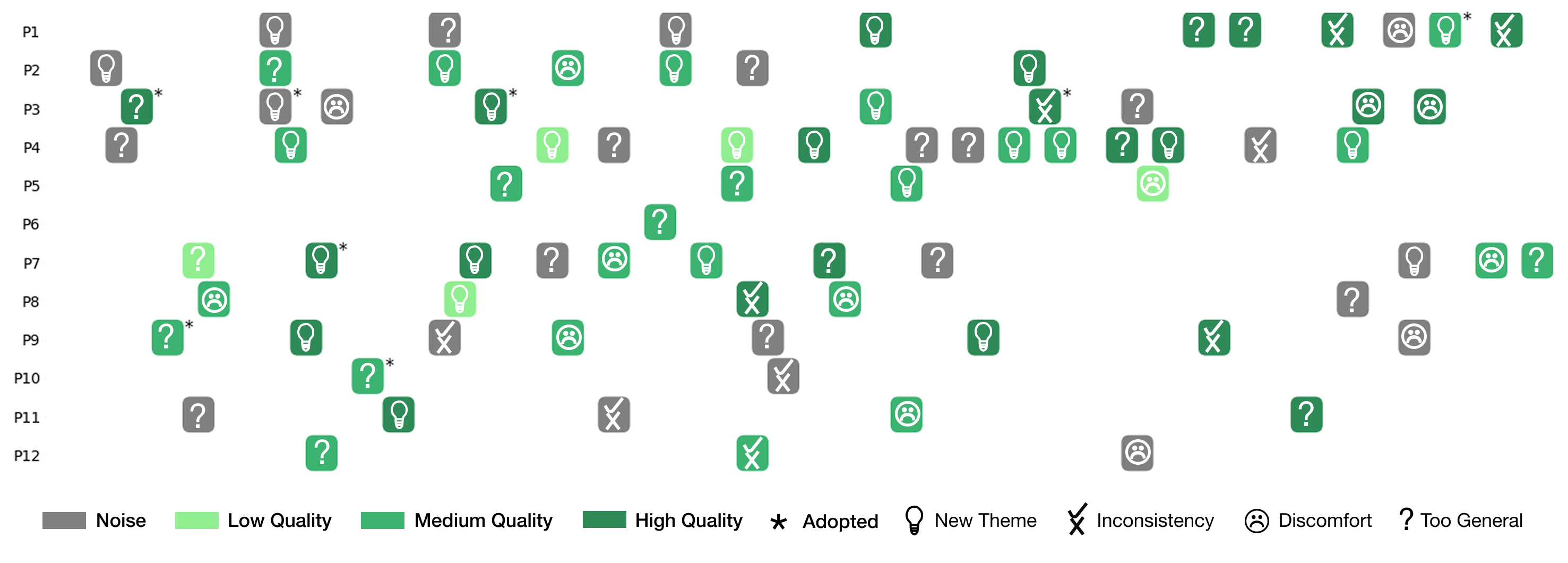}
    \caption{Distribution of \system{}’s suggestions over the course of the interview. The figure visualizes each suggestion’s type, expert-rated quality, and whether it was adopted by the interviewer.}
    \Description{.}
    \label{suggestionSchedule}
\end{figure*}

\subsection{How well does \system{} support interviewers' real-time data sensemaking?}
\subsubsection{Capturing Desired Information}
Participants preferred \system{} over the baseline for active information capture (Baseline: M = 2.42, SD = 1.08; \system{}: M = 5.67, SD = 0.89; \textit{p} < .001, \textit{d} = .002). On average, they created 4.8 manual tags (SD = 3.3) and clicked the AI summarization button 9.1 times (SD = 2.7), suggesting complementary use of manual and AI-assisted tags. Manual tags were typically short phrases rather than long notes, often reflecting interviewers’ interpretations or conclusions. For instance, when an interviewee described Netflix’s auto play scenario, P8 created the tag ``never continue watching'', explaining that details mattered less than the conclusion. Many manual tags also captured non-verbal information not present in transcripts, such as facial expressions or gestures—for example, P5 tagged ``froned'' to note an interviewee’s reaction.

In contrast, AI-assisted tags were used to condense complex factual information or when interviewers could not quickly summarize a response. For example, for the question ``What other things are you doing when you are watching?'', the AI generated ``usually focused on the show unless distracted.'' Interviewers also relied on AI when they only partially understood an answer but did not want to interrupt. As P7 described, ``When I realize what the interviewee just said is valuable but I can’t summarize it in a short phrase right away, I click summarize to let the AI help. Sometimes I also use it when I didn’t fully understand the response, but wanted to ensure my understanding.'' The ``Click to summarize'' feature was rated positively (M = 5.17, SD = 1.89), with participants noting that AI summaries often captured exactly what they wanted to record, see \autoref{fig:rating-features}. They also valued having control over when to invoke AI, contrasting it with fully automated summaries, which they felt produced generic overviews and risked obscuring important details.

\subsubsection{Surfacing interview content worth further exploration}
\label{qualitative results of follow-up suggestions}
\h{
Participants preferred \system{} for suggestions on potentially overlooked content and follow-up opportunities.

Participants considered the suggestions easy to understand and poses minimal cognitive overhead. ``The frequency of suggestions is appropriate, which didn't distract me a lot, and the concise format with visual icons allows me to check the system using my peripheral vision,'' noted P5. However, they held mixed opinions about the utilities of the suggestions from \system{}. Some participants described suggestions they received as ``unexpectedly accurate,'' noting that they faithfully reflected the underlying interaction context. For example, when an interviewee vaguely responded ``I would erase my personal data when necessary,'' the system flagged ``no specific steps described,'' which the interviewer found helpful in prompting deeper inquiry. Some participants felt that certain suggestions added little value because they highlighted observations that were already noticed, such as tagging repeated fillers with ``hesitation, discomfort.'' While some participants felt these cues increased their burden, others appreciated them as an additional perspective that boosted their confidence. As P10 noted, ``Sometimes the system just confirms what I already saw, but that confirmation makes me more certain that I should probe further.'' Overall, participants felt that suggestions about newly emerged themes or inconsistencies often contained information that they missed, whereas cues about emotional discomfort or overly general answers tended to be readily noticeable to interviewers. 
  
\autoref{suggestionSchedule} showed the distribution of \system{}'s suggestions over the whole interview session, together with experts' average ratings on the quality of the suggestions and users' adoption behaviors. Combining user ratings (M = 4.50, SD = 1.24, see \autoref{fig:rating-features}) with expert evaluations in \autoref{tab:system-performance}, we observe a divergence: 
Although experts rated the baseline system’s suggestions as relatively higher quality on average (\system{}: M=2.21, SD= 0.59 Baseline: M= 2.66, SD=0.35), users did not prefer them in practice. We attribute this discrepancy to the time-sensitive nature of live interviewing. Suggestions from the baseline system were often lengthy and interviewers did not have the cognitive bandwidth to verify during the fast-paced flow of the interview. They reported that the suggestions in the baseline were difficult to parse and required considerable time to read. Although these suggestions were often more actionable, typically phrased as direct questions that interviewers could ask, participants did not view this actionability as desirable. 
As P3, who adopted a baseline suggestion, explained: ``Cuz I just saw it and at that moment I was sort of blanking out, so I just asked. But then the interviewee looked confused, and it didn’t go anywhere.''

From the interaction logs, we found that \system{} generated 6.5 (SD = 3.9) tags about situations where interviewers often need to explore further per interview, with 10.2\% led to a follow-up question compared with 4.83 (SD = 3.12) suggestions from the baseline and a 15.5\% adoption rate, which means less than one suggestion was adopted for each interview in both conditions, on average. Each participant asked 24.58 (SD = 6.42) questions on average using \system{}, and 18.17 (SD = 7.09) using baseline. This suggests that the actionability of suggestions is limited. As P3 said, ``It makes sense that the system flagged this as an inconsistency, but what should I do? I don’t know how to naturally shift the focus to that point, even if the system reminds me that a clarification probe should be used here.'' In rapidly evolving interview conversations, many participants prioritize maintaining a natural conversational flow over pursuing every potential follow-up opportunity, even when they recognize that such follow-ups could yield additional insights.
}

\begin{figure*}[!ht]
    \centering
    \includegraphics[width=\linewidth]{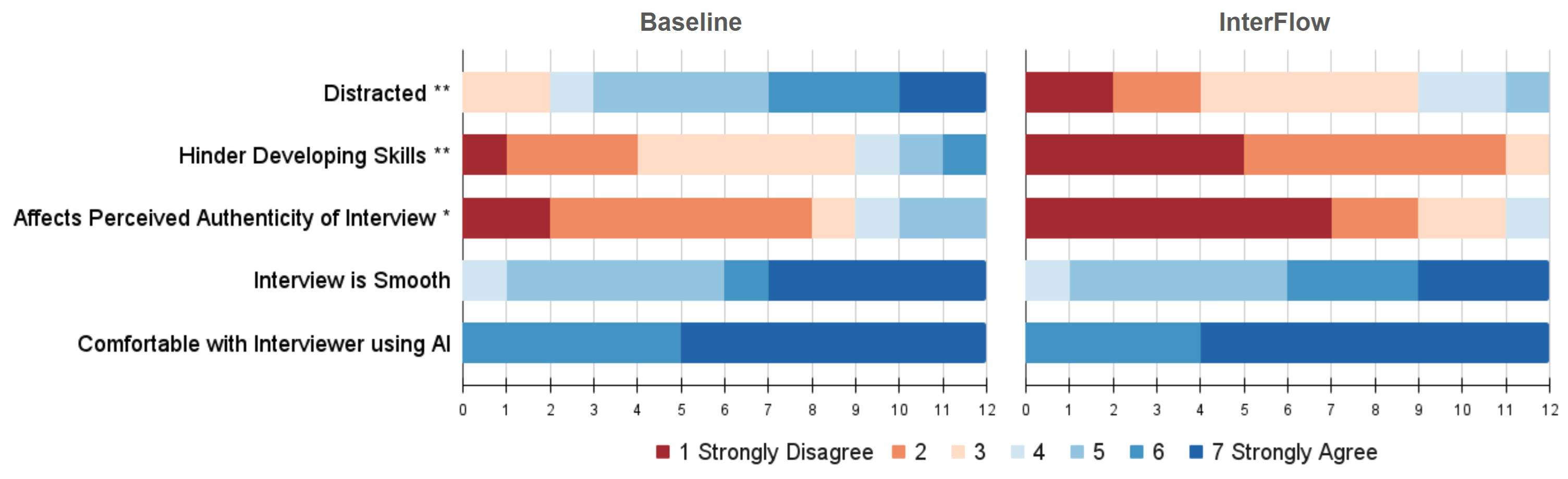}
    \caption{Distribution of users' perceptions of Potential Risks ( $p < 0.05$ is marked with *, and $p < 0.01$ is marked with **) }
    \Description{.}
    \label{rating of risks}
\end{figure*}

\subsection{Concerns and Potential Risks}
Prior research on AI support in online meetings~\cite{areWeOnTrack} and clinical decision-making~\cite{unremarkable} highlights risks that often arise in cognitively demanding, time-pressured settings. Building on these insights, we investigated whether similar concerns emerge when using \system{}. Specifically, we examined participants’ perceptions of risks related to (1) distraction from the interview conversation, (2) reduced authenticity of the interview, (3) hindrance to the interviewer’s skill development, and (4) negative impacts on the interviewee’s experience.

\subsubsection{Distraction from the main activity}
Participants reported that \system{} was significantly less distracting than the baseline (Baseline: M = 5.17, SD = 1.34; \system{}: M = 2.83, SD = 1.19; \textit{p} = .007, \textit{d} = 1.07). Although not distraction-free, they felt the benefits outweighed the cost. As P6 explained, ``Of course the system takes a little bit of my attention, but the help it provides makes this level of distraction acceptable.''

Compared with the baseline, participants emphasized that \system{}’s design reduced the need to constantly monitor the screen. They attributed this to clear, unobtrusive visualizations, concise and interpretable suggestions, flexibility in when to engage, and well-calibrated notifications. As P3 noted, ``I didn’t need to keep staring at the interface; the AI’s summaries were never more than one sentence, and the suggestions combined a simple visual icon with a short excerpt, making them intuitive to understand.''

Some remote participants raised a related concern: shifting attention to the interface reduced their ability to notice facial expressions and subtle gestures. While not considered a severe drawback, they viewed it as a trade-off—a slight reduction in non-verbal awareness in exchange for stronger support in managing the interview.

\subsubsection{Perceived authenticity of the interview}
We examined perceived authenticity and found a significant difference (Baseline: M = 2.58, SD = 1.38; \system{}: M = 1.75, SD = 1.06; \textit{p} = .047, \textit{d} = 0.75). In the baseline condition, participants often requested follow-up questions but seldom used them, as the assistant’s fully formed questions felt prescriptive and reduced interviewer agency. In contrast, \system{} offered lightweight tags with icons and excerpts that were optional and non-intrusive. As P7 noted, ``It never felt like the system was speaking for me. The suggestions were in the background, and I still had to decide how to use them.'' Overall, \system{} better preserved authenticity by scaffolding interviewer judgment and maintaining a natural interaction flow.



\subsubsection{Skill development}

Results showed a clear difference in concerns about skill development (Baseline: M = 3.08, SD = 1.38; \system{}: M = 1.67, SD = 0.65; \textit{p} = .010, \textit{d} = 1.03). Participants worried more that the baseline system might limit their opportunity to practice core interviewing abilities, whereas they generally did not feel this way about \system{}.

Qualitative feedback explained this gap. With the baseline assistant, some interviewers adopted its follow-up questions without reflection, raising concerns about bypassing the learning process. As P5 put it, ``When I just asked the system’s question, it felt like skipping the thinking process. If I did that too often, I wouldn’t get better at asking my own.'' In contrast, \system{}’s lightweight tags acted as scaffolds rather than replacements, prompting interviewers to interpret and phrase their own follow-ups. As P8 noted, ``The system reminded me of things I could ask, but I still had to think about the exact wording and timing. It feels like I’m practicing, not outsourcing.'' 

Overall, \system{} preserved interviewer agency and was perceived as more supportive of long-term skill development.

\subsubsection{Influence on interviewee's experience}
Interviewees generally did not report much discomfort with the interviewer’s use of AI (Baseline: M = 6.58, SD = 0.51; \system{}: M = 6.67, SD = 0.49; \textit{p} = 1.000, \textit{d} = -0.16). Some noted feeling slightly less attended to but framed this as an inevitable part of technology’s growing presence. As one interviewee said, ``It felt like when you are talking to someone and they get momentarily distracted by a message on their phone—there is a slight sense of discomfort, but it is something quite common in daily life.'' Others linked their comfort to the non-sensitive nature of the topics. ``Maybe because the topic we were talking about is not that sensitive, I’m okay with AI being part of the conversation.''

No significant differences in perceived smoothness were found (Baseline: M = 5.83, SD = 1.11; \system{}: M = 5.67, SD = 0.98; \textit{p} = .750, \textit{d} = 0.20), though interviewees acknowledged subtle effects. In particular, they observed occasional pauses as interviewers processed system suggestions, which slightly interrupted the conversational flow.

Overall, these findings reveal an asymmetry: the system eased interviewers’ cognitive load but still left interviewees aware of minor disruptions in engagement and smoothness.

\section{Discussion}

\subsection{Design Implications for AI Assistance under Time-constrained, Cognitively Demanding Scenarios }
\h{\subsubsection{Design Process-oriented AI Assistance that can Fit into User's Thinking Flow}
Unlike collaborating with a human co-interviewer, where shared research context, mutual understanding, and comparable capabilities are largely assumed—AI’s limitations and reliability remain opaque to humans. As a result, interviewers cannot fully trust it and must pause to evaluate them before acting. Surfacing AI's ``unfinished thoughts'', like the situation tag in \system{}, improves the transparency of its reasoning and provides a ``shared cognitive space''~\cite{liu2025thought} between user and AI. When AI's output fits naturally into interviewers' thinking flow, they can make sense of the AI's suggestions with minimal effort, thus truly benefiting from AI assistance. As suggested by Liu et. al ~\cite{liu2025thought}, the “thought process” of AI should not be viewed solely as a mechanism for explainability, but also as a “substrate for human–AI interaction”. Such a process-centered approach may extend to other high-stakes domains, including clinical decision making~\cite{unremarkable} and classroom settings~\cite{zhang2025cpvis}.

At the same time, such partial cues enabled users to recognize divergences between their intentions and AI's suggestions, identify potential errors, and be resilient~\cite{glassman2024ai} to AI's outputs. This is critical for reducing under- and over-reliance on AI~\cite{swaroop2024accuracy}. In such exploratory knowledge-intensive tasks, fully delegating control over conversational direction to AI risks de-skilling human beings and may pose broader challenges to intellectual development ~\cite{ibrahim2025measuring}.

\subsubsection{Fine-grained and Dynamic User Modeling}

Our user study reveals that even among interviewers with comparable levels of expertise, perceptions of the utility of AI suggestions might vary. This variability suggests that effective AI assistance requires more fine-grained personalization to align with individual user’s preferences. Importantly, many of these preferences—such as how frequently users want AI to intervene—are difficult to articulate in advance or to specify through natural language. Future agentic systems should learn these implicit user preferences through users' accumulated interaction traces (i.e. whether suggestions are viewed, ignored, or acted upon in our case) and refine its behavior ~\cite{borghoff2025human}. 

At the same time, our study demonstrates the benefits of proactive AI assistance in reducing users’ effort and enabling them to meaningfully process AI suggestions, even when they are primarily focused on their main task. However, in rapidly evolving conversational settings, the utility of such suggestions is highly contingent on whether they align with the current conversational dynamics and whether users have sufficient cognitive bandwidth to engage with them.

In our current system, we rely on voice activity as a single signal to determine when suggestions should be presented. We assumed that a lack of voice activity may indicate a moment of reflection for the interviewer, and thus a potentially appropriate time to surface suggestions. Yet a single signal alone is limited and may not reliably reflect the user's cognitive state, as recent work on proactive assistance for cooking tasks ~\cite{huh2025vid2coach} has noted. By identifying additional indicators of user attention and developing more advanced methods to estimate them accurately, proactive assistance can be achieved with a lower risk of disturbance and greater usefulness.

}
\subsection{Design Natural Interaction Modalities to Support the Triangular Dynamic in AI-assisted Human-human Activity}

Semi-structured interviews are inherently triadic: interviewer, interviewee, and AI assistant. Participants noted that even subtle gaze shifts to the system could reduce their sensitivity to non-verbal cues, potentially making interviewees feel less attended to. Similar triangular dynamics are likely to emerge in other human–human communication settings where AI mediates interaction, such as AI-facilitated clinical consultations ~\cite{unremarkable} or AI-enhanced teacher–student interactions\cite{tang2024vizgroup,zhang2025cpvis}. Beyond sensing users' context with intelligent agents to reduce the requirement for explicit user input, designing more natural interaction modalities with devices that better integrate into the human body, such as smart glasses or smart watches, can further minimize noticeable attention shifts, allowing users to receive assistance without disrupting social engagement.

\h{\subsection{Limitations and Future Work}

\subsubsection{Evaluation Setting}
Our current evaluation employed an in-lab study to provide an initial assessment of the system’s usability and usefulness in supporting semi-structured interviews. Although we tried to approximate realistic interview conditions (see \autoref{efforts for enhancing realism}), this setting still differs from how interviewers would use the system in their own interviews, where they are more familiar with the script, have more knowledge about the topic, and stronger motivation to conduct a high-quality interview. How real-world usage and long-term adoption might affect users’ perceptions of the system remains unknown.

\subsubsection{Data Privacy}
Our current implementation relies on the OpenAI Realtime API, which uses proprietary, closed-source models that cannot be locally deployed. As a result, data privacy cannot be fully guaranteed—an issue that is particularly critical in qualitative interviews, where sensitive personal information is often discussed. To move toward in-the-wild usage, future work should find alternative solutions with open-source models capable of performing real-time, complex speech understanding. 

\subsubsection{System Robustness}
The system’s robustness remains limited. Our experiment shows that using off-the-shelf text-embedding models to encode the interview conversation and script questions does not achieve high accuracy in detecting the ongoing question(0.58 on average across all participants). Future work should explore more advanced information retrieval methods(e.g. learning-based) that can reliably retrieve the corresponding scripted question even when interviewers change the wording. Also, the suggestions generated by \system{} contain a noticeable amount of noise. Future work could explore how to better manage conversational context to reduce hallucinations, as well as how to better identify and filter out suggestions with factual error.
}
\subsubsection{Broader Applicability}
We see opportunities to extend \system{} beyond semi-structured interviews to other cognitively demanding, time-sensitive tasks such as usability testing, clinical consultations, or educational tutoring. These domains share similar challenges of balancing active listening, task management, and real-time sensemaking, and could benefit from adaptive scaffolds like those explored in this work.

\section{Conclusion}
In this paper, we present \system{}, a novel interactive system that supports interviewers in managing interview flow and facilitates real-time sensemaking of interview data. The system’s ambient visualizations and mixed-initiative design impose minimal cognitive overhead and avoid distracting interviewers from their primary conversation with the interviewee. While the system’s multi-level, process-oriented suggestions integrate well with interviewers’ thinking flow, their actionability is limited by the complex and rapidly evolving conversational dynamics of semi-structured interviews. Future work should explore more advanced approaches to modeling users and their surrounding contexts in order to provide more personalized AI assistance in attention-intensive and time-constrained scenarios.

\begin{acks}
The first author started this project as her final year project at City University of Hong Kong. We thank Dr. Sangho Suh, Xinyue Chen, and Dr. Chengbo Zheng for their insightful feedback on this project. We also thank Dr. Ping Ma and Dr. Ananya Tiwari for validating and enriching the scheme for suggesting follow-up questions in semi-structured interviews.
\end{acks}
\bibliographystyle{ACM-Reference-Format}
\bibliography{sample-base}

\clearpage

\appendix

\setlength{\intextsep}{5pt}

\section{Prompt Templates}
In this section, we provide the prompt templates used in \system{}. Python-style string interpolation variables (e.g., \$\{variable\}) represent dynamic content inserted at runtime.

\subsection{Prompt Template for Analyzing Script Structure}
This prompt is used to parse the raw text of the interview script into a hierarchical JSON structure for visualization.

\begin{lstlisting}[style=promptstyle]
Analyze the given interview script and extract question categories, the main questions for each category, and sub-questions under each main question.

Return the result in the following JSON format:
[
  {
    "category": "Stage 1",
    "questions": [
      {
        "main_question": "",
        "sub_questions": [
          { "sub_question": "" },
          { "sub_question": "" }
        ]
      },
      {
        "main_question": "",
        "sub_questions": [
          { "sub_question": "" }
        ]
      }
    ]
  },
  {
    "category": "Stage 2",
    "questions": [
      {
        "main_question": "",
        "sub_questions": []
      }
    ]
  }
]
Make sure the output is valid JSON. Extract the intro that provides context or background for the questions in each category.
\end{lstlisting}

\subsection{Prompt Template for Instant Summary}
This prompt is triggered when the user clicks to summarize a recent segment of the conversation. We utilized the \textbf{Claude 3.5 Sonnet} model for this feature to ensure rapid summarization generation.

\begin{lstlisting}[style=promptstyle]
This is the transcript of an interview, please infer who is the interviewee, and then give me an extractive summary of what the interviewee said, that answers the interviewer's question.

Reply with no more than seven words, do not include meaningless words such as "interviewee describes..."
\end{lstlisting}

\subsection{Prompt Template for Detecting Situations Worth Further Exploration}
This prompt runs continuously to identify potential probe or follow-up opportunities based on the conversation history. We utilized the \textbf{GPT-4o Realtime Preview} (version 2025-06-03) model to minimize latency during the live interview session.

\begin{lstlisting}[style=promptstyle]
You are an observer of a semi-structured interview regarding the research question: ${globalResearchQuestion} and the background: ${globalBackground}.

Your ONLY job is to return the situation where a follow up or probe is needed, following the rules listed below.

Cases that needs probe:
To manage the conversation, ask for elaboration, detail. Keep the interview on target. Ask for clarification, examples, evidence.
1.1 Help reveal slant or bias. The response is unclear or too general, e.g. "It just feels better that way." Or when the interviewee refers to shared knowledge, or uses unclear pronouns, gestures or jargon e.g. "you know how it is", "everyone does that".
1.2 When the interviewee hesitates, self-corrects, feels down, which could signal discomfort, deeper meaning. e.g. "Well... I mean... not exactly..."

Cases that needs follow up:
To get depth, detail, richness, vividness and nuance, helping to assure thoroughness and credibility. Explore relevant events, concepts, and themes. Designed in response to the comments or ideas introduced by the conversational partner.
2.1 New concept or theme emerged that is relevant to the research question ${globalResearchQuestion}, especially when the interviewee said a substitute word or relevant term. e.g. interviewer asked about user experience for the car seat in the interviewee's own car, but interviewee started to talk about car seat problem in school bus.
2.2 When there is apparent contradiction or inconsistency in what the interviewee has said. e.g. "I don't think ads bother me." / "Sometimes ads are so annoying..." or "You told me before that the hats were green, but just now you referred to them as blue. Are they sometimes green and sometimes blue, or do they just look blue to you sometimes?"

Please return the summarized content of conversation you detected that matches the situation in no more than one sentence, and then give me the number of the situation.

The output should be in the following format:
{
  "situation": "The specific content in the conversation",
  "number": "The number of the situation(e.g. 1.1, 1.2, 2.1, 2.2)"
}
\end{lstlisting}

\subsection{Prompt Template for Generating Further Suggestions}
Once a situation is detected, this prompt generates specific strategies for the interviewer (displayed when hovering over a tag).

\begin{lstlisting}[style=promptstyle]
/* If it is a probe situation: */
Based on the situation detected, please check what kind of probe is suitable based on the conversational context: ask for clarification, or ask for evidence and example.

/* If it is a follow-up situation: */
Based on the situation detected, please check what kind of follow-up strategy is suitable based on the conversational context:

Strategies about follow-up on concepts:
What to follow up on: 
1. Negated concepts or concepts with negative meanings
2. Comments, attitudes, or emotional expressions
General Strategies: 
1. Repeat or paraphrase to confirm 
2. Request a definition
POS-based Strategies: 
1. Reduce the scope of a noun concept 
2. Probe details about a verb concept 
3. Probe the degree or the reason of an adjective or adverb concept

Strategies about general follow-ups:
- Why? Request potential causes of a result or reasons for a comment, an opinion, or an expressed emotion
- Can you be more specific? Request for elaboration or details when the response PQ is too short or little specific information is provided
- And then? Request a complete timeline
- For example? Request an example for a general noun such as "thing", "content", or "object"
- What else? Ask for extra supplements when the response PQ is already adequate

Strategies for Follow-ups on Related Concepts:
General Strategies: 
1. Compare between concepts
2. Clarify whether A equals B 
3. Bring up another concept of the same class
4. Bring up a hypothesis or an example

POS-based Strategies: 
1. When the selected concept is a noun:
   - bring up its usage or capabilities
   - bring up one of its attributes
   - bring up a subclass or superclass
   - request the respondent's comments on it in a certain aspect
2. When the selected concept is a verb:
   - bring up a specific timing when it happens
   - bring up a place or platform where it happens
   - bring up a tool related to it
   - bring up a certain way or degree it happens in
   - bring up its cause or consequence
   - bring up its precondition
3. When the selected concept is an adjective or an adverb:
   - bring up its superlative degree
\end{lstlisting}

\subsection{Prompt Template for Evaluating Detected Situations}
This prompt is used to evaluate the collection of situations detected in a sliding time window to filter out low-quality suggestions. We utilized the \textbf{GPT-4o} model for this task to leverage its higher reasoning capabilities for quality assurance.

\begin{lstlisting}[style=promptstyle]
There are several situations detected during a part of a semi-structured interview that worth following up or probing. Please rate each situation with 1-5 based on the criteria described below. The rationale is as follows:

Criteria

Correctness
Definition: Whether the detected situation is supported by the evidence from the interview transcript.
How to apply: Check if the detected situation faithfully reflects what the interviewee actually said or implied. Incorrect or hallucinated ones would fail this criterion.

Specificity
Definition: Whether the detected situation is precise and clearly distinguishable, as opposed to vague or overly general.
How to apply: Look at whether the detected situation pinpoints a concrete idea, behavior, or inconsistency, rather than repeating generic concepts (e.g., "privacy issue" vs. "no specific steps described for erasing personal data").

Coverage
Definition: Whether the detected situation captures obvious or important points without omitting critical events in the conversation.
How to apply: Examine whether the system overlooked major responses that human coders would reasonably expect to notice.
\end{lstlisting}

\section{Participants}
The demographic information of participants in our observational study and user evaluation is presented in \autoref{tab:observational-demographics} and \autoref{tab:evaluation-demographics}, and interviewee demographics of user evaluation is presented in \autoref{tab:evaluation-demographics-interviewee}.


\begin{table*}[!ht]
  \centering
  \caption{Demographic information of observational study participants and their expertise level for semi-structured interviews.}
  \label{tab:observational-demographics}
  \footnotesize
  \setlength{\tabcolsep}{4pt}
  \renewcommand{\arraystretch}{1.2}

  \begin{tabularx}{0.9\textwidth}{l l l c c >{\raggedright\arraybackslash}p{4.5cm} l}
    \toprule
    \textbf{No.} & \textbf{Gender} & \textbf{Age} & \textbf{Field of Study} & \textbf{Current Position} & \textbf{Number of Times Conducting Interview} & \textbf{Expertise Level} \\
    \midrule
    P1 & Female & 21 & HCI & Undergrad Research Assistant & $\,(5,10]$ & 2 \\
    P2 & Male & 20 & HCI & Undergrad Research Assistant & $\,(0,5]$ & 2 \\
    P3 & Female & 25 & Sociology & Master Student & $\,(10,15]$ & 4 \\
    P4 & Male & 25 & HCI & PhD Student & $\,(10,15]$ & 4 \\
    P5 & Female & 26 & HCI & PhD Student & $\,(15,20]$ & 5 \\
    P6 & Female & 25 & HCI & PhD Student & $\,(5,10]$ & 3 \\
    \bottomrule
  \end{tabularx}

  \vspace{4pt}
  \begin{minipage}{\textwidth}
    \footnotesize
    For the number of times conducting semi-structured interviews, we define one instance as a single 1-on-1 interview session. The self-rated expertise level ranges from 1 (beginner) to 7 (expert).
  \end{minipage}
\end{table*}


\begin{table*}[!ht]
  \centering
  \caption{Demographic information of evaluation study participants who acted as interviewers and their expertise level for semi-structured interviews.}
  \label{tab:evaluation-demographics}
  \footnotesize
  \setlength{\tabcolsep}{4pt}
  \renewcommand{\arraystretch}{1.2}

  \begin{tabularx}{0.9\textwidth}{l l l c c >{\raggedright\arraybackslash}p{4.2cm} l}
    \toprule
    \textbf{No.} & \textbf{Gender} & \textbf{Age} & \textbf{Field of Study} & \textbf{Current Position} & \textbf{Number of Times Conducting Interview} & \textbf{Expertise Level} \\
    \midrule
    P1  & Female & 39 & Sociology & PhD Student & $\,(0,5]$      & 2 \\
    P2  & Male   & 21 & HCI & Undergrad Research Assistant & $\,(0,5]$      & 1 \\
    P3  & Female & 26 & HCI & Product Designer            & $\,(20,\infty)$ & 6 \\
    P4  & Female & 20 & Psychology & Master Student        & $\,(10,15]$    & 5 \\
    P5  & Female & 27 & HCI & Master Student              & $\,(15,20]$    & 5 \\
    P6  & Female & Prefer Not to Say & HCI & PhD Student    & $\,(15,20]$    & 5 \\
    P7  & Female & 22 & HCI & PhD Student                 & $\,(5,10]$     & 4 \\
    P8  & Male   & 25 & Privacy and Security & PhD Student & $\,(0,5]$      & 1 \\
    P9  & Female & Prefer Not to Say & HCI & PhD Student   & $\,(20,\infty)$ & 5 \\
    P10 & Female & 26 & Economics & PhD Student           & $\,(5,10]$     & 4 \\
    P11 & Female & 29 & Public Health & PhD Student       & $\,(20,\infty)$ & 5 \\
    P12 & Male   & 25 & HCI & PhD Student                & $\,(0,5]$      & 2 \\
    \bottomrule
  \end{tabularx}

  \vspace{4pt}
  \begin{minipage}{\textwidth}
    \footnotesize
    For the number of times conducting semi-structured interviews, we define one instance as a single 1-on-1 interview session. The self-rated expertise level ranges from 1 (beginner) to 7 (expert).
  \end{minipage}
\end{table*}




\begin{table*}[h]
  \centering
  \caption{Demographic information of evaluation study participants who acted as interviewee.}
  \label{tab:evaluation-demographics-interviewee}
  \footnotesize
  \setlength{\tabcolsep}{4pt}
  \renewcommand{\arraystretch}{1.2}
\begin{tabular}{l l l}
    \toprule
    \textbf{No.} & \textbf{Gender} & \textbf{Age} \\
    \midrule
    P1  & Female & 23 \\
    P2  & Male   & 26 \\
    P3  & Male & 27 \\
    P4  & Female & 23 \\
    P5  & Female & 25 \\
    P6  & Male & 26 \\
    P7  & Female & 25 \\
    P8  & Female   & 26 \\
    P9  & Prefer Not to Say & Prefer Not to Say \\
    P10 & Male & 31 \\
    P11 & Female & 19 \\
    P12 & Female   & 29 \\
    \bottomrule
\end{tabular}
  \vspace{4pt}
\end{table*}


\end{document}